\documentstyle[12pt]{article}

\makeatletter
\def\section{\@startsection{section}{1}{\z@}{-3.5ex plus -1ex minus -.2ex}
{2.3ex plus .2ex}{\large\bf}}
\makeatother

\makeatletter
\@addtoreset{equation}{section}

\makeatother

\parskip=\medskipamount

\def\qed{\vrule height 1.5ex width 1.2ex depth -.1ex}

\begin{document}

\input epsf

\vspace{8mm}

\begin{center}

{\Large \bf R--torsion and linking numbers from simplicial abelian gauge
theories} \\

\vspace{12mm}

{\large David H. Adams}

\vspace{4mm}

School of Mathematics, Trinity College, Dublin 2, Ireland. \\

\vspace{1ex}

email: dadams@maths.tcd.ie

\end{center}

\begin{abstract}

Simplicial versions of topological abelian gauge theories are constructed
which reproduce the continuum expressions for the partition function and
Wilson expectation value of linked loops, expressible in terms of 
R--torsion and linking numbers respectively. The new feature which makes 
this possible is the introduction of simplicial fields (cochains) 
associated with the dual triangulation of the background
manifold, as well as with
the triangulation itself. This doubling of fields, reminiscent of lattice
fermion doubling, is required because the natural simplicial analogue of
the Hodge star operator maps between cochains of a triangulation and 
cochains of the dual triangulation. The simplicial analogue of 
Hodge--de Rham theory is developed, along with a natural simplicial
framework for considering linking numbers of framed loops.
When the loops represent torsion elements of the homology of the 
manifold then ${\bf Q}/{\bf Z}$--valued torsion pairings appear in place of 
linking numbers for certain discrete values of the coupling parameter of 
the theory.

\end{abstract}

\newpage

\section{Introduction}

In this paper we consider the problem of discretising abelian
Chern--Simons gauge theory on closed 3-manifolds,
and its generalisation to arbitrary odd-dimensional closed manifolds,
in such a way that the topological objects of interest in the theories
are equal to the corresponding objects in the discrete theories. 
The topological objects of interest are the partition function and
Wilson vacuum expectation values (v.e.v.'s) of linked loops. 
As is well-known, the partition functions (suitably gauge-fixed and 
analytically regularised) can be expressed in terms of analytic Ray--Singer
torsion \cite{Schwarz}, while the Wilson v.e.v.'s of linked loops (and
their generalisations to higher dimensions) can be expressed in terms of 
the linking numbers of the loops 
\cite{Poly} \cite[\S2]{W(Jones)}.\footnote{These properties have been
checked in \cite{CM}, where the abelian Chern--Simons theory on $S^3$
was explicitly solved.}
The Ray--Singer torsion \cite{RS} of a manifold has a combinatorial analogue,
namely the Reidemeister--Franz torsion \cite{Reid,Franz,Milnor,RS},
also known as the R--torsion, defined via a 
triangulation of the manifold. The cochain complex of the triangulation
(with values in ${\bf R}\,$, or more generally in a flat vector bundle)
plays a role in the construction of the R--torsion which is analogous
to the role of the de Rham complex in the construction of the Ray--Singer
torsion. These torsions are in fact known to be
equal \cite{Muller,Cheeger}.
This leads us to consider discrete versions of abelian Chern--Simons theory
(and its generalisations) where the fields are cochains
of a triangulation of the background manifold. The aim is to find such a
theory where the partition function is expressible in terms of R--torsion
(in place of Ray--Singer torsion), and the Wilson v.e.v.'s of linked 
loops are expressible in terms of the linking numbers, as in the continuum 
theory.

Discrete versions of abelian Chern--Simons theory 
have been considered previously by a number of authors \cite{ES,Pullin},
including setups where the gauge field is taken to be a cochain of a 
triangulation \cite{KM,BR(PLB),Albeverio}, although it seems that no
explicit attempt has been made previously to carry out the aims outlined 
above\footnote{A way of
obtaining linking numbers from a lattice version of abelian Chern-Simons
theory is discussed in the recent preprint ref. \cite{Pullin}, although
the setup and techniques used there are quite different from ours.}.
The new feature of our approach, which allows to carry out these aims, is
the introduction of simplicial fields (cochains) associated with the
{\it dual} of the triangulation, as well as with the triangulation 
itself\footnote{The dual triangulation has appeared previously in connection
with discrete abelian ${\bf Z}_p$ field theories, although not in the way that 
we describe here; see \cite{Rakowski(PR)} and the ref.'s therein. 
Duality has previously played a key role in a variety of lattice 
field theories, see e.g. \cite{Cardy,Frohlich}.}.
The motivation behind this is intimately connected with the nature of the
simplicial analogue of the Hodge star operator.
In the continuum abelian Chern--Simons theory the Hodge star operator, and
its interrelationship with the exterior derivative (and the adjoint of the
exterior derivative), play a crucial role in deriving the expressions 
for the partition function and Wilson v.e.v.'s of loops. 
If the same evaluation of the partition function in a cochain version of the 
theory is to lead to an analogous expression with R--torsion in place of
Ray--Singer torsion then there must be a simplicial analogue of the 
Hodge star operator such that its interrelationship with the cochain action 
functional and cochain derivative (simplicial analogue of the exterior 
derivative) are the same as in the continuum case.
In fact there is a natural simplicial analogue of the 
Hodge star operator, namely the duality operator mapping between 
cochains of the triangulation and cochains of the dual of the 
triangulation. (The cochains we are considering are over the reals.)
After taking into account the fact that this operator maps
between different cochain complexes, its interrelationship with the derivative
map for cochains of the triangulation and the derivative map for cochains
of the dual triangulation is completely analogous to the interrelationships
in the continuum case. (We will see this explicitly in proposition 2.4.)
This indicates that if a cochain version of abelian Chern--Simons theory
is to have its partition function expressible in terms of R--torsion in the
desired way then the theory should involve cochains of the dual triangulation
as well as of the triangulation itself. This is reminiscent of the field
doubling in lattice fermion theories (see e.g. \cite{BeJo}).

Motivated by these considerations we consider discretisation of the double
abelian Chern--Simons theory (and its generalisation to higher dimensions)
with two independent gauge fields, such that in the discrete version one
gauge field is a cochain of the triangulation and the other is a cochain
of the dual triangulation. We show that, after making a certain ``twist''
in the action functional which couples the fields but leaves the expressions
for the partition function and Wilson v.e.v.'s of loops 
unchanged\footnote{except for the removal of a metric-dependent phase in
the partition function}, 
there is a canonical discretisation (i.e. cochain
version) of this theory such that the interrelationship between the 
discrete (i.e. simplicial) action functional, 
the simplicial Hodge star operator
(i.e. the duality operator) and the cochain derivatives is the same as in the 
continuum case. As a consequence the evaluation of the partition function
gives an expression involving R--torsion in the desired way, reproducing
the continuum expression\footnote{The continuum expression is reproduced
when the value of the coupling parameter is $\lambda\!=\!1$. For 
$\lambda\ne1$ a triangulation-dependent renormalisation of $\lambda$ is
required to reproduce the continuum expression.}.
We consider Wilson v.e.v.'s of framed linked 
loops (and their generalisation to higher dimensions) which fit into this
discrete setup in a natural way: The framed loops are taken to be 
ribbons with one boundary being an edge 
loop in the triangulation and the other boundary being an edge 
loop in the dual triangulation. (We call these framed loops {\it simplicial
framings} of edge loops.)
We find that the Wilson v.e.v.'s of these linked, simplicially framed 
loops can be expressed in terms of the linking
numbers of the loops, again reproducing the continuum 
expression\footnote{This is true for all values of the coupling parameter
$\lambda$ --no triangulation-dependent renormalisation is required in this 
case.}.

The key tool in constructing the discretisation is the {\em Whitney map}
\cite{Whitney}, a canonical map from cochains to differential forms on the
background manifold\footnote{This map played a key role in establishing the
equality between R--torsion and Ray--Singer torsion \cite{Dodziuk,Muller}.}.
This map has previously been used to construct simplicial versions of
quantum field theories in \cite{AlZe}, and was used to construct a 
simplicial version of the abelian Chern--Simons theory in 
\cite[\S5]{Albeverio}. In both cases the emphasis was on proving convergence
of the simplicial theory to the continuum limit --the problem of reproducing
topological quantities was not addressed. There is a crucial difference
between the way the Whitney map is used in \cite{AlZe,Albeverio} and the 
way we use it here: We embed the cochains of the triangulation and the 
cochains of its dual into the space of cochains of the {\em barycentric
subdivision} of the triangulation, and then use the Whitney map on the cochains
of the barycentric subdivision to obtain differential forms on the background
manifold. (In \cite{AlZe,Albeverio} the Whitney map was used directly 
on the cochains of the triangulation --the dual triangulation and barycentric
subdivision did not enter.)

It turns out that the doubled, ``twisted'' abelian Chern--Simons theory 
which we discretise is identical to the abelian BF--theory for two 
independent gauge fields. Thus we end up with a canonical discrete version
of abelian BF--theory for which the partition function and Wilson v.e.v.'s of 
linked framed loops are expressed in terms of R--torsion and linking
numbers respectively, and are equal to the corresponding objects in the 
continuum theory. (Note that this equality does not require a continuum
limit --it holds exactly for each triangulation of the background manifold).

The linking numbers $\mbox{lk}(\gamma^{(1)},\gamma^{(2)})$ of bounding loops 
$\gamma^{(1)}\,$, $\gamma^{(2)}$ enter in the expressions for the 
Wilson v.e.v.'s of loops through factors of the form 
\begin{eqnarray*}
\exp\left(\frac{i\pi^2}{\lambda}
n_1n_2\mbox{lk}(\gamma^{(1)},\gamma^{(2)})\right)
\end{eqnarray*}
where $\lambda$ is the coupling
parameter of the theory and $n_1$ and $n_2$ are integers. These expressions
are trivial when the coupling parameter takes the discete values
$\lambda=\frac{\pi}{2p}\,$, $p\in{\bf Z}$. 
We show that non-trivial expressions for the Wilson
v.e.v.'s can be obtained at these discrete values of $\lambda$
in the more general case
where the loops may represent torsion elements of the 
${\bf Z}-$homology of the manifold. (Bounding loops represent the zero-element
in this homology.) There is a canonical ${\bf Q}/{\bf Z}-$valued pairing
between torsion elements (of appropriate degree) in the ${\bf Z}-$homology
of $M\,$, and we show (proposition 6.1) that the pairing of torsion elements
represented by loops can be expressed as a generalisation of linking number.
As a consequence ${\bf Q}/{\bf Z}-$valued torsion pairings appear
in place of ${\bf Z}-$valued
linking numbers in the expressions for the Wilson v.e.v.'s in this case.

The mathematical results which we require are derived in \S2--3.
In \S2 the simplicial analogue of Hodge--de Rham theory is developed,
culminating in the main result, theorem 2.6, which gives a natural
simplicial analogue of the following basic formula:
\begin{eqnarray}
\int_Md\omega\wedge\tau=<{\ast}d\omega\,,\,\tau>
\label{1.1}
\end{eqnarray}
where $\omega$ and $\tau$ are differential forms on a closed manifold $M$
and the inner product $<\cdot\,,\cdot>$ and Hodge star operator $\ast$
are determined by a metric on $M$. In \S3 a natural simplicial framework
for considering linking numbers of framed loops is developed. 
The first main result, theorem 3.3,
gives a formula for the linking number of an edge loop 
in a triangulation $K$ and an edge loop in the 
dual triangulation $\widehat{K}$.
More generally, this result is a formula for the linking number
$\mbox{lk}(f_K,g_{\widehat{K}})$ of a simplicial map $f_K$ 
from a triangulated closed manifold of dimension $p$ and a dual-simplicial map 
$g_{\widehat{K}}$ from a triangulated closed manifold
of dimension $q$ into a triangulated manifold $M$ of dimension $n=p+q+1$.
A particular class of framings of edge loops --namely the simplicial 
framings described above-- is introduced, 
and it is proved that such framings always exist
(theorem 3.6). This is used in proposition 3.9 to give a formula for the 
linking number of edge loops in $K$ in the simplicial
setup: The linking number is expressed in terms of disjoint
simplicial framings of the loops, and is shown to be independent of the choice
of framings.
(This is the formula which appears in the expression for the Wilson v.e.v.'s
of linked framed loops in the discrete abelian gauge theory).
In \S4 we review the formal evaluation of the partition function and 
Wilson v.e.v.'s of 
loops in abelian Chern--Simons theory and its associated BF theory
(including features associated with the moduli space of flat $U(1)$
gauge fields which appear when $\pi_1(M)$ is non-trivial). In \S5 we apply
the results of \S2--3 to construct the simplicial version of the abelian
BF gauge theory described above, and show that it produces the R--torsion
and linking numbers as claimed, leading to agreement with the continuum
expressions for the partition function and v.e.v.'s obtained in \S4.
A group of simplicial gauge transformations is introduced which enables 
the features of the continuum theory involving the modulispace of flat
$U(1)$ gauge fields to also be reproduced in the simplicial theory.
In \S6 we consider Wilson v.e.v.'s of loops representing torsion elements
of the homology of $M\,$, and show that in this case for certain discrete 
values of the coupling parameter the v.e.v. is a purely homological quantity,
expressible in terms of torsion pairings (as described above). 

Background on the mathematical objects and constructions appearing in this 
paper can be found for example in \cite{Novikov} \cite{Munkres}.

The techniques and results of this paper are of a general nature and
we expect them to have application in a variety contexts.
These may include lattice fermion doubling
\cite{BeJo}, a new approach to lattice gauge theory based on non-commutative
geometry\cite{Balachandran}\footnote{The possible 
applicability of our work in this 
context was pointed out by Prof. A.P.~Balachandran in a discussion.},
developing discrete versions of theories for
topologically massive (abelian) gauge fields
\cite{Jackiw}, anyons and high $T$ superconductivity \cite{Wil,Poly},
quantum gravity in the loop variable approach\footnote{Linking numbers 
have recently appeared in this context \cite{Ashtekar} and work aiming 
at a lattice formulation is also in progress \cite{Pullin}.}
and reproducing the S--duality in abelian gauge theory demonstrated by Witten 
\cite{W(Sdual)}\footnote{I thank Prof. G.~Segal for pointing out this
reference.}. Since abelian Chern--Simons theory can be considered as the weak
coupling limit of the non-abelian theory this paper may provide a pointer
to how to discretise the non-abelian Chern--Simons theory, or rather
its associated BF theory. The aim in this case would be to
reproduce the combinatorial
3-manifold invariant of Turaev and Viro \cite{TV} from the partition function, 
and the (generalised) Jones polynomial and related knot invariants from the 
Wilson v.e.v.'s (see \cite{CottaRamusino} for a discussion of these in the
continuum case).

\section{Simplicial analogue of Hodge--de Rham theory}

In this section we develop a simplicial analogue of Hodge--de Rham theory
in which the simplicial analogue of the Hodge star operator is the duality
operator between cochains of a triangulation and cochains of the dual
triangulation (over the reals).

Let $M$ be a smooth closed oriented manifold of dimension $n\,$, and let
$K$ be a simplicial complex smoothly triangulating $M$. If $\{v_0,v_1,\dots,
v_p\}$ are the vertices of a $p$-simplex in $K$ we denote the oriented
$p$-simplex with orientation specified by the ordering $v_0,\dots,v_p$ by
$\alpha=\alpha^{(p)}={\lbrack}v_0,\dots,v_p\rbrack\,$, and denote the 
closed cell in $M$ corresponding to $\alpha$ by $|\alpha|$. 
We write $\alpha^{(p)}<\beta^{(q)}$ to denote that the $p$-simplex 
$\alpha^{(p)}$ is a face in the $q$-simplex $\beta^{(q)}$ ($p<q$).
The vectorspace $C_p(K)$ is the space of formal finite linear 
combinations (``$p$-chains'') of oriented
$p$-simplices of $K$ over the reals with the rule that, as an element of
$C_p(K)\,$, an oriented $p$-simplex changes sign under a change of 
orientation, i.e. ${\lbrack}v_{\tau(0)},\dots,v_{\tau(p)}\rbrack=(-1)^{\tau}
{\lbrack}v_0,\dots,v_p\rbrack$ for permutation $\tau$ of $\{0,\dots,p\}$.
The vectorspace $C^p(K)$ of $p$-cochains is the dual of $C_p(K)$. 
There is a canonical inner product in $C_p(K)$ determined by requiring that
the oriented $p$-simplices be orthonormal; this gives an identification
$C_p(K)\,{\cong}\,C^p(K)$ and we will consider oriented $p$-simplices 
as elements of $C^p(K)$ as well as $C_p(K)$.
The boundary operator $\partial^{K}:C_p(K){\to}C_{p-1}(K)$ is the linear
operator which maps an oriented $p$-simplex $\alpha^{(p)}$ to the sum of 
its $(p-1)$-faces with orientations induced by 
the orientation of $\alpha^{(p)}$.
The coboundary operator $d^{K}:C^p(K){\to}C^{p+1}(K)$ is the adjoint of
$\partial^{K}\,$; thus $d^{K}{\lbrack}v_0,\dots,v_p\rbrack=\sum_v{\lbrack}
v,v_0,\dots,v_p\rbrack$ where the sum is over all vertices $v$ such
that ${\lbrack}v,v_0,\dots,v_p\rbrack$ is a $(p+1)$-simplex.
These operators have the property $\partial^{K}\partial^{K}=0$ and
$d^{K}d^{K}=0$.

The star $\mbox{St}(\alpha)$ of a $p$-simplex $\alpha$ is the open 
region in $M$ consisting of the union of the interiors
of all simplices (of any degree) which contain $\alpha$ as a face.
Recall that the barycentric coordinate function $\mu_v:M\to{\bf R}$ 
corresponding to a 
vertex $v$ of $K$ is the function equal to one at $v{\in}M\,$, decreasing
linearly (in a suitable sense) to zero at the boundary of $\mbox{St}(v)$ 
and vanishing outside $\mbox{St}(v)$.
It has the properties
\begin{eqnarray}
\sum_v\mu_v=1\qquad\quad\mbox{and}\qquad\quad\sum_{v\in\alpha}\mu_v(y)=1
\quad{\forall}y\in|\alpha|\;\;\;\forall\alpha{\in}K
\label{2.1}
\end{eqnarray}
Let $\Omega^p(M)$ denote the vectorspace of ${\bf R}$--valued $p$-forms on
$M$ which are smooth on the complement of a set of measure zero, and let
$d:\Omega^p(M)\to\Omega^{p+1}(M)$ denote the exterior derivative.
The {\it Whitney map} $W^{K}:C^p(K)\to\Omega^p(M)$ is the linear map 
defined by \cite{Whitney} \cite{Dodziuk}
\begin{eqnarray}
W^{K}(\alpha)&=&p!\sum_{i=0}^p(-1)^i
\mu_id\mu_0\wedge\cdots{\wedge}d\mu_{i-1}{\wedge}
d\mu_{i+1}\wedge\cdots{\wedge}d\mu_p\quad\mbox{for}\quad
\alpha={\lbrack}v_0,\dots,v_p\rbrack \nonumber \\
& &\label{2.2}
\end{eqnarray}
($W^{K}(v)=\mu_v$) where $\mu_i:=\mu_{v_i}$. The functions $\mu_v$ are 
smooth on the complement of the $(n-1)$-skeleton of $K\,$, a set of measure
zero in $M\,$, so the same is true of $W^{K}(\alpha)$. Clearly $W^{K}$ 
vanishes outside of $\mbox{St}(\alpha)$ and on the boundary of
$\mbox{St}(\alpha)$.
Recall that the {\it de Rham map} $\,A^{K}:\Omega^p(M){\to}C^p(K)$ is 
the linear map 
defined by $<A^{K}(\omega)\,,\alpha>=\int_{|\alpha|}\omega$ for each 
oriented $p$-simplex $\alpha{\in}K$. The maps $W^{K}$ and $A^{K}$ have
the properties
\begin{eqnarray}
A^{K}W^{K}=\mbox{Id}\ \ \mbox{(the identity)}\qquad\quad&,&\qquad
\int_{|\beta|}W^{K}(\alpha)=<\alpha\,,\beta> \label{2.3} \\
d\,W^{K}=W^{K}d^{K}\qquad\qquad&,&\qquad\qquad\,d^{K}A^{K}=A^{K}d \label{2.4}
\end{eqnarray}
(see \cite{Whitney} \cite{Dodziuk}). It follows from (\ref{2.3}) that
$W^{K}$ is injective.

{\it Simplicial wedge product.} We define the bilinear product 
$\wedge^{K}:C^p(K){\times}C^q(K){\to}$ $C^{p+q}(K)$ by
\begin{eqnarray}
x\wedge^{K}y:=A^{K}(W^{K}(x){\wedge}W^{K}(y))
\label{2.5}
\end{eqnarray}
The following properties follow easily from (\ref{2.3})--(\ref{2.4}):
\hfill\break
(i) Skewsymmetry: $x\wedge^{K}y=(-1)^{pq}y\wedge^{K}x$ \hfill\break
(ii) Leibniz rule: $d^{K}(x\wedge^{K}y)=d^{K}x\wedge^{K}y+(-1)^p
x\wedge^{K}d^{K}y$ \hfill\break
However, as we will see below, $\wedge^{K}$ is not associative.
This product, with $K$ replaced by its barycentric subdivision $BK\,$, 
will play a key role in proving our results in the following. It has an
explicit description given in the proposition below, which will allow
us to translate differential-geometric problems into combinatorial ones.

{\it Convention 2.1.} If $\{v_0,\dots,v_p\}$ 
are vertices in $K$ but are not
the vertices of a $p$-simplex then we set ${\lbrack}v_0,\dots,v_p\rbrack:=0$ 
in $C^p(K)$.

\noindent With this convention we have 

{\it Proposition 2.2.} ${\lbrack}v_0,\dots,v_p\rbrack\wedge^{K}
{\lbrack}w_0,\dots,w_q\rbrack=0$ if the vertices $\{v_0,\dots,v_p\}$
and $\{w_0,\dots,w_q\}$ do not have precisely one element in common, and
\begin{eqnarray}
{\lbrack}v_0,\dots,v_p\rbrack\wedge^{K}
{\lbrack}v_p,\dots,v_{p+q}\rbrack=\frac{p!q!}{(p+q+1)!}{\lbrack}
v_0,\dots,v_{p+q}\rbrack\,.
\label{2.6}
\end{eqnarray}

\noindent {\it Proof.} 
The proposition obviously holds if either $\alpha={\lbrack}v_0,\dots,v_p
\rbrack$ or $\beta={\lbrack}w_0,\dots,w_q\rbrack$ are not simplices in $K$.
Assume that $\alpha$ and $\beta$ are simplices.
If the sets $\{v_0,\dots,v_p\}$ and $\{w_0,\dots,w_q\}$
have no common elements, or have one element in common but their union
is not the vertices of a $(p+q)$-simplex, then 
$\mbox{St}(\alpha)\cap\mbox{St}(\beta)$ is disjoint from all $(p+q)$-simplices
of $K\,$, so the restriction of $W^{K}(\alpha){\wedge}W^{K}(\beta)$
to any $(p+q)$-simplex of $K$ (considered as a region in $M$) vanishes, 
hence $\alpha\wedge^{K}\beta=0$.
If the sets have two or more elements in common then $W^{K}(\alpha)
{\wedge}W^{K}(\beta)$ is a sum of terms with each term containing a factor
$d\mu_i{\wedge}d\mu_i$ where $\mu_i$ is the barycentric coordinate function
of one of the shared vertices. This vanishes, hence $\alpha\wedge^{K}\beta=0$. 
When $\beta={\lbrack}v_p,\dots,v_{p+q}\rbrack$ the only possibility for a 
$(p+q)$-simplex meeting $\mbox{St}(\alpha)\cap\mbox{St}(\beta)$ is
${\lbrack}v_0,\dots,v_{p+q}\rbrack\,$, so
${\lbrack}v_0,\dots,v_p\rbrack\wedge^{K}{\lbrack}v_p,\dots,v_{p+q}\rbrack$
is proportional to ${\lbrack}v_0,\dots,v_{p+q}\rbrack$. To complete the
proof we calculate
\begin{eqnarray*}
\lefteqn{<{\lbrack}v_0,\dots,v_p\rbrack\wedge^{K}{\lbrack}v_p,\dots,v_{p+q}
\rbrack\,,\,{\lbrack}v_0,\dots,v_{p+q}\rbrack>} \\
& &\qquad=\int_{|{\lbrack}v_0,\dots,v_{p+q}\rbrack|}W^{K}({\lbrack}v_0,
\dots,v_p\rbrack){\wedge}W^{K}({\lbrack}v_p,\dots,v_{p+q}\rbrack) \\
& &\qquad=\int_{\left\{ {\scriptstyle0{\le}\mu_i{\le}1\ \ ,\ i=1,\dots,p+q 
\atop\scriptstyle0{\le}\mu_1+\dots+\mu_{p+q}{\le}1}\right\}}p!q!\mu_p
d\mu_1\wedge\cdots{\wedge}d\mu_{p+q} \\
& &\qquad=\frac{p!q!}{(p+q+1)!}\qquad\qquad\qquad\qed
\end{eqnarray*}

The proposition shows that $\wedge^{K}$ is an ``antisymmetrisation'' of 
the usual cup product. In this form it has appeared previously 
in the literature
in connection with simplicial discretisation of abelian Chern-Simons theory
\cite{BR(PLB)} \cite{Albeverio}. It also follows that $\wedge^{K}$ is
non-associative: For $x{\in}C^p(K)\;,\;y{\in}C^q(K)\;,\;z{\in}C^r(K)$
\begin{eqnarray}
x\wedge^{K}(y\wedge^{K}z)=\Bigl(\frac{p+q+1}{r+q+1}\Bigr)\,(x\wedge^{K}y)
\wedge^{K}z
\label{2.7}
\end{eqnarray}
is an easy consequence of (\ref{2.6}).

Let $BK$ denote the barycentric subdivision of $K$ and let $\widehat{K}$ 
denote the dual triangulation, i.e. the cell decomposition of $M$ dual to $K$.
These will play central roles in the following, and can be characterised as 
follows. The vertices of $BK$ are $\{\widetilde{\alpha}\,|\,\alpha{\in}
K\}$ where $\widetilde{\alpha}$ denotes the barycenter of the simplex 
$\alpha{\in}K$. The oriented $p$-simplices of $BK$ are
\begin{eqnarray*}
\{{\lbrack}\widetilde{\alpha}^{(q_0)},\dots,
\widetilde{\alpha}^{(q_p)}\rbrack\,|\,
\alpha^{(q_0)}<\alpha^{(q_1)}<\cdots<\alpha^{(q_p)}{\in}K\}.
\end{eqnarray*}
For each oriented p-simplex $\alpha^{(p)}{\in}K$ we denote the 
oriented $(n-p)$-cell 
in $M$ dual to $\alpha^{(p)}$ by $\widehat{\alpha^{(p)}}$. It is the 
cell corresponding to the union
of all $(n-p)$-simplices in $BK$ of the form ${\lbrack}\tilde{\alpha}^{(p)},
\dots,\tilde{\alpha}^{(n)}\rbrack$ where $\alpha^{(p)}<\alpha^{(p+1)}<
\cdots<\alpha^{(n)}{\in}K$. The orientation of $\widehat{\alpha^{(p)}}$ is 
uniquely determined by the orientations of $\alpha^{(p)}$
and $M$ as follows. If $\alpha^{(0)}<\alpha^{(1)}<\cdots<\alpha^{(p)}<
\cdots<\alpha^{(n)}{\in}K$ is such that ${\lbrack}\tilde{\alpha}^{(0)},\dots,
\tilde{\alpha}^{(p)}\rbrack$ is compatible with the orientation of 
$\alpha^{(p)}$ and ${\lbrack}\tilde{\alpha}^{(0)},\dots,\tilde{\alpha}^{(n)}
\rbrack$ is compatible with the orientation of $M$ then the orientation of
$\widehat{\alpha^{(p)}}$ is compatible with that of 
${\lbrack}\tilde{\alpha}^{(p)},\dots,\tilde{\alpha}^{(n)}\rbrack$.
The dual triangulation is the cell decomposition of $M$ given by
$\widehat{K}=\{\widehat{\alpha}\,|\,\alpha{\in}K\}$.

The complexes $\{C_*(BK)\,,\,\partial^{BK}\}\,$, $\{C^*(BK)\,,\,d^{BK}\}$
and $\{C_*(\widehat{K})\,,\,\partial^{\widehat{K}}\}\,$, $\{C^*(\widehat{K})
\,,\,d^{\widehat{K}}\}$ are defined analogously to 
$\{C_*(K)\,,\,\partial^K\}\,$, $\{C^*(K)\,,\,d^K\}$.
We define injective linear maps
\begin{eqnarray}
B:C^p(K){\hookrightarrow}C^p(BK)\qquad\quad,\qquad\quad\,B:C^p(\widehat{K})
{\hookrightarrow}C^p(BK)
\label{2.8}
\end{eqnarray}
as follows. For oriented $p$-simplex $\alpha^{(p)}{\in}K$ we set 
\begin{eqnarray}
B\alpha^{(p)}:=\frac{1}{N(\alpha^{(p)})}\sum\limits_{\alpha^{(0)}<\alpha^{(1)}<
\dots<\alpha^{(p)}}\pm{\lbrack}\tilde{\alpha}^{(0)},\tilde{\alpha}^{(1)},
\dots,\tilde{\alpha}^{(p)}\rbrack
\label{2.9}
\end{eqnarray}
where $N(\alpha^{(p)})\;(=p!)$ is the number of terms in the sum and the sign
$\pm$ is chosen so that $\pm{\lbrack}\tilde{\alpha}^{(0)},\dots,\tilde
{\alpha}^{(p)}\rbrack$ has orientation compatible with $\alpha^{(p)}\,$,
and set
\begin{eqnarray}
B\widehat{\alpha^{(p)}}:=
\frac{1}{N(\widehat{\alpha^{(p)}})}\sum\limits_{\alpha^{(p)}<\alpha^{(p+1)}<
\dots<\alpha^{(n)}}\pm{\lbrack}\tilde{\alpha}^{(p)},\tilde{\alpha}^{(p+1)},
\dots,\tilde{\alpha}^{(n)}\rbrack
\label{2.10}
\end{eqnarray}
where $N(\widehat{\alpha^{(p)}})$ is the number of terms in the sum 
and the sign $\pm$ is chosen so that $\pm{\lbrack}\tilde{\alpha}^{(p)},
\dots,\tilde{\alpha}^{(n)}\rbrack$ has orientation compatible with 
$\widehat{\alpha^{(p)}}$. 
These maps are orthogonality-preserving, i.e. $<x_1,x_2>=0$ implies
$<Bx_1,Bx_2>=0\,$, and the images of $C^p(K)$ and $C^p(\widehat{K})$ under $B$
are orthogonal in $C^p(BK)$.

{\it Simplicial Hodge star operator.} We define the linear operators
$\ast^K:C^p(K){\to}C^{n-p}(\widehat{K})$ and $\ast^{\widehat{K}}:
C^q(\widehat{K}){\to}C^{n-q}(K)$ by 
\begin{eqnarray}
<\ast^Kx\,,y>&:=&\frac{(n+1)!}{p!(n-p)!}\int_MW^{BK}(Bx){\wedge}W^{BK}(By)
\label{2.11} \\
<\ast^{\widehat{K}}y\,,x>&:=&\frac{(n+1)!}{p!(n-p)!}\int_MW^{BK}(By)
{\wedge}W^{BK}(Bx) \label{2.12} 
\end{eqnarray}
for $x{\in}C^p(K)\,$, $y{\in}C^{n-p}(\widehat{K})$. These definitions are 
analogous to the definition of the Hodge star operator on differential forms
(modulo the numerical factors).
The operator $\ast^K$ coincides with the usual duality operator:

{\it Proposition 2.3.} Let $\alpha^{(p)}{\in}K$ be an oriented
$p$-simplex with dual cell $\widehat{\alpha^{(p)}}\in\widehat{K}$. Then
\begin{eqnarray*}
\ast^K\,\alpha^{(p)}=\widehat{\alpha^{(p)}}\qquad\mbox{and}\qquad
\ast^{\widehat{K}}\,(\widehat{\alpha^{(p)}})=(-1)^{p(n+1)}\alpha^{(p)}\,.
\end{eqnarray*}

\noindent {\it Proof.} Using (\ref{2.4}) the operator $\ast^K$ can be 
expressed in terms of the simplicial wedge product for $C^*(BK)\,$:
\begin{eqnarray}
<\ast^Kx\,,y>=\frac{(n+1)!}{p!(n-p)!}\,<(Bx)\wedge^{BK}(By)\,,\,{\lbrack}M
\rbrack_{BK}>
\label{2.13}
\end{eqnarray}
where ${\lbrack}M\rbrack_{BK}$ is the orientation $n$-cocycle of $M$ in
$C^n(BK)\,$, i.e. the sum of all $n$-simplices in $BK$ with orientations
induced by that of $M$. The proposition is now an easy consequence of
proposition 2.2 (with $K$ replaced by $BK$)
together with the characterisations of $BK$ and $\widehat{K}$
given above and the definition of the maps $B$. \qed

{\it Proposition 2.4.} 
\begin{eqnarray*}
&\mbox{(i)}&\quad(\ast^K)^*=(\ast^K)^{-1}\qquad\qquad\mbox{and}
\qquad\qquad(\ast^{\widehat{K}})^*=(\ast^{\widehat{K}})^{-1}   \\
&\mbox{(ii)}&\quad\ast_{n-p}^{\widehat{K}}\,\ast_p^K=(-1)^{p(n+1)}\mbox{Id}
\qquad\ \ \mbox{and}\ \ \qquad\ast_{n-q}^K\,\ast_q^{\widehat{K}}=(-1)^{q(n+1)}
\mbox{Id} \\
&\mbox{(iii)}&\ \ (d_p^K)^*=(-1)^{np+1}\ast_{n-p}^{\widehat{K}}d_{n-p-1}^
{\widehat{K}}\ast_{p+1}^K\quad\mbox{and}\quad(d_q^{\widehat{K}})^*=
(-1)^{nq+1}\ast_{n-q}^Kd_{n-q-1}^K\ast_{q+1}^{\widehat{K}} 
\end{eqnarray*}

\noindent {\it Proof.} Parts (i) and (ii) are immediate consequences of
proposition 2.3. To prove (iii) we derive a formula for $(d^{\widehat{K}})^*=
\partial^{\widehat{K}}$. For $\alpha^{(p)}={\lbrack}v_0,\dots,v_p\rbrack{\in}
K$ using proposition 2.3 we find
\begin{eqnarray}
\partial^{\widehat{K}}\ast^K\alpha^{(p)}&=&\partial^{\widehat{K}}\widehat
{\alpha^{(p)}}=\sum_{\alpha^{(p+1)}>\alpha^{(p)}}\widehat{\alpha^{(p+1)}}=
\ast^K\left(\sum_{\alpha^{(p+1)}>\alpha^{(p)}}\alpha^{(p+1)}\right)\nonumber\\
&=&\ast^K\left(\sum_{v_{p+1}}{\lbrack}v_0,\dots,v_p,v_{p+1}\rbrack\right)=
(-1)^{p+1}\ast^K\left(\sum_{v_{p+1}}{\lbrack}v_{p+1},v_0,\dots,v_p\rbrack
\right) \nonumber \\
&=&(-1)^{p+1}\ast^Kd^K\alpha^{(p)} \label{2.14}
\end{eqnarray}
i.e. $(d_{n-p-1}^{\widehat{K}})^*\ast_p^K=\partial_{n-p}^{\widehat{K}}
\ast_p^K=\ast_{p+1}^Kd_p^K$. Part (iii) follows easily from this formula
and (ii) and (i). \qed

\noindent After taking into account the fact that $\ast^K$ and 
$\ast^{\widehat{K}}$ map between different cochain complexes the formulae
in proposition 2.4 are completely analogous to the formulae relating the 
Hodge star operator on differential forms to the exterior derivative and its
adjoint. A simple consequence of this proposition is Poincare duality: the
operator $\ast^K$ provides an isomorphism between the cohomology spaces
of $d^K$ and $d^{\widehat{K}}$. Of course, this is already well-known since
$\ast^K$ coincides with the usual duality operator. The novel aspect of the
preceding is that we have seen (in proposition 2.3) how $\ast^K$ arises
as a simplicial analogue of the Hodge star operator, which in turn shows 
how Poincare duality arises as the simplicial analogue of Hodge duality
(via proposition 2.4). In contrast, the standard construction of the duality
operator (see e.g. \cite[\S67]{Munkres}) is via the cup and cap products,
which do not have natural analogues in Hodge--de Rham theory. (For example,
cup and cap products require an ordering of the vertices of $K$).

Our aim in the remainder of this section is to derive a simplicial analogue
of the basic Hodge--de Rham theoretic formula (\ref{1.1}). 
This is theorem 2.6 below, which will allow
us to construct a simplicial action functional with the desired properties 
in \S5.

{\it Proposition 2.5.} Consider
\begin{eqnarray}
<d^K{\lbrack}v_0,\dots,v_p\rbrack\wedge^K{\lbrack}w_{p+1},\dots,w_n\rbrack\,,\,
{\lbrack}M\rbrack_K>
\label{*}
\end{eqnarray}
where ${\lbrack}M\rbrack_K$ is the orientation $n$-cocycle of $M$ in $C^n(K)$.
This equals $\pm(-1)^{p+1}\left\lbrack{n+1\atop{p+1}}\right\rbrack^
{(-1)}$ if ${\lbrack}v_0,\dots,v_p,w_{p+1},\dots,w_n\rbrack$ is an 
$n$-simplex in $K\,$, with sign $\pm$ determined by its orientation relative
to $M\,$, and vanishes otherwise.

\noindent {\it Proof.} Since $d^K{\lbrack}v_0,\dots,v_p\rbrack=\sum_{v_{p+1}}
{\lbrack}v_{p+1},v_0,\dots,v_p\rbrack$ we see from proposition 2.2 that
(\ref{*}) is potentially non-vanishing in the following two cases:
(i) $\{v_0,\dots,v_p\}$ and $\{w_{p+1},\dots,w_n\}$ have precisely one element
in common, or (ii) $\{v_0,\dots,v_p,w_{p+1},\dots,w_n\}$ are the vertices of 
an n-simplex in $K$. We must show that (\ref{*}) vanishes in case (i). In this
case we can assume without loss of generality that
${\lbrack}w_{p+1},\dots,w_n\rbrack={\lbrack}v_p,v_{p+1},\dots,v_{p-1}\rbrack$.
Then, by proposition 2.2, 
\begin{eqnarray*}
d^K{\lbrack}v_0,\dots,v_p\rbrack\wedge^K{\lbrack}v_p,\dots,v_{n-1}\rbrack
&=&\sum_w{\lbrack}w,v_0,\dots,v_p\rbrack\wedge^K{\lbrack}v_p,\dots,v_{n-1}
\rbrack \\
&=&\frac{(p+1)!(n-p-1)!}{(n+1)!}\sum_w{\lbrack}w,v_0,\dots,v_{n-1}\rbrack
\end{eqnarray*}
and
\begin{eqnarray*}
{\lbrack}v_0,\dots,v_p\rbrack\wedge^Kd^K{\lbrack}v_p,\dots,v_{n-1}\rbrack
&=&\sum_w{\lbrack}v_0,\dots,v_p\rbrack\wedge^K{\lbrack}w,v_p,\dots,v_{n-1}
\rbrack \\
&=&(-1)^p\frac{(p+1)!(n-p-1)!}{(n+1)!}\sum_w{\lbrack}w,v_0,\dots,v_{n-1}\rbrack
\end{eqnarray*}
so that, using the Leibniz rule, we get
\begin{eqnarray*}
d^K{\lbrack}v_0,\dots,v_p\rbrack\wedge^K{\lbrack}v_p,\dots,v_{n-1}\rbrack
=\frac{1}{2}d^K({\lbrack}v_0,\dots,v_p\rbrack\wedge^K{\lbrack}v_p,
\dots,v_{n-1}\rbrack)\,.
\end{eqnarray*}
It follows that 
\begin{eqnarray*}
(\ref{*})=\frac{1}{2}<{\lbrack}v_0,\dots,v_p\rbrack\wedge^K{\lbrack}v_p,
\dots,v_{n-1}\rbrack\,,\,(d^K)^*{\lbrack}M\rbrack_K>=0
\end{eqnarray*}
since $(d^K)^*{\lbrack}M\rbrack_K=\partial^K{\lbrack}M\rbrack_K=0$.
To complete the proof we calculate (\ref{*}) in case (ii):
\begin{eqnarray*}
\lefteqn{
<d^K{\lbrack}v_0,\dots,v_p\rbrack\wedge^K{\lbrack}v_{p+1},\dots,v_n\rbrack\,,\,
{\lbrack}M\rbrack_K>} \\
&=&\sum_{k=1}^{n-p}<{\lbrack}v_{p+k},v_0,\dots,v_p\rbrack\wedge^K{\lbrack}
v_{p+1},\dots,v_n\rbrack\,,\,{\lbrack}M\rbrack_K> \\
&=&\sum_{k=1}^{n-p}(-1)^{p+1}\frac{(p+1)!(n-p-1)!}{(n+1)!}<{\lbrack}v_0,\dots,
v_n\rbrack\,,\,{\lbrack}M\rbrack_K> \\
&=&\pm(-1)^{p+1}\frac{(p+1)!(n-p)!}{(n+1)!}\qquad\quad\qed
\end{eqnarray*}

We define the bilinear functional $S:C^p(K){\times}C^{n-p-1}(\widehat{K})
\to{\bf R}$ by
\begin{eqnarray}
S(x,y)&=&\left\lbrack{\scriptstyle{n+1}\atop\scriptstyle{p+1}}\right\rbrack
\int_MdW^{BK}(Bx){\wedge}W^{BK}(By)
\quad,\quad(x,y){\in}C^p(K){\times}C^{n-p-1}(\widehat{K}) \nonumber \\
& &\label{2.15}
\end{eqnarray}
where $\left\lbrack{n+1\atop{p+1}}\right\rbrack=
\frac{(n+1)!}{(p+1)!(n-p-1)!}\,$,
and define the linear maps $T_p^K:C^p(K){\to}C^{n-p-1}(\widehat{K})$
and $T_{n-p-1}^{\widehat{K}}=(T_p^K)^*:C^{n-p-1}(\widehat{K}){\to}
C^p(K)$ by
\begin{eqnarray}
S(x,y)=<T_p^Kx\,,\,y>=<x\,,\,T_{n-p-1}^{\widehat{K}}y>\,.
\label{2.16}
\end{eqnarray}

{\it Theorem 2.6.}
\begin{eqnarray*}
T^K=\ast^Kd^K\qquad\quad\mbox{and}\qquad{\quad}T_{n-p-1}^{\widehat{K}}=
(-1)^{np+1}\ast^{\widehat{K}}d^{\widehat{K}}
\end{eqnarray*}

{\it Corollary 2.7.}
\begin{eqnarray*}
T^{\widehat{K}}T^K=(d^K)^*d^K\qquad\quad\mbox{and}\qquad{\quad}
T^KT^{\widehat{K}}=(d^{\widehat{K}})^*d^{\widehat{K}}
\end{eqnarray*}

\noindent {\it Proof.} We must show that
\begin{eqnarray}
\left\lbrack{\scriptstyle{n+1}\atop\scriptstyle{p+1}}\right\rbrack
\int_MdW^{BK}(B\alpha^{(p)}){\wedge}W^{BK}
(B(\widehat{\beta^{(p+1)}}))=<\ast^Kd^K\alpha^{(p)}\,,\,
\widehat{\beta^{(p+1)}}>
\label{2.17}
\end{eqnarray}
for arbitrary oriented $p$-simplex $\alpha^{(p)}$ and $(p+1)$-simplex
$\beta^{(p+1)}$ in $K$.
The theorem and its corollary then follow easily from proposition 2.4.
Using (\ref{2.4}) the l.h.s. of (\ref{2.17}) can be expressed in terms of the 
simplicial wedge product for $C^*(BK)\,$:
\begin{eqnarray*}
\mbox{l.h.s.}=
\left\lbrack{\scriptstyle{n+1}\atop\scriptstyle{p+1}}\right\rbrack
<d^{BK}(B\alpha^{(p)})\wedge^{BK}
(B\widehat{\beta^{(p+1)}})\,,\,{\lbrack}M\rbrack_{BK}>
\end{eqnarray*}
It follows from proposition 2.5 (with $K$ replaced by $BK$) that for sequences
of simplices $\alpha^{(0)}<\dots<\alpha^{(p)}{\in}K$ and $\beta^{p+1}<\dots<
\beta^{(n)}{\in}K$ the quantity
\begin{eqnarray*}
<d^{BK}{\lbrack}\tilde{\alpha}^{(0)},\dots,\tilde{\alpha}^{(p)}\rbrack
\wedge^{BK}{\lbrack}\tilde{\beta}^{(p+1)},\dots,\tilde{\beta}^{(n)}\rbrack
\,,\,{\lbrack}M\rbrack_{BK}>
\end{eqnarray*}
vanishes unless $\alpha^{(p)}<\beta^{(p+1)}\,$, and is equal to 
$\pm(-1)^{p+1}\left({n+1\atop{p+1}}\right)^{-1}$ in the latter case, so 
from the definition of the maps $B$ we get
\begin{eqnarray}
\mbox{l.h.s.}=\left\{{\pm(-1)^{p+1}\qquad\qquad\quad\mbox{for}\quad
\alpha^{(p)}<\beta^{(p+1)}\atop{0\qquad\qquad\quad\mbox{otherwise}}}\right.
\label{2.18}
\end{eqnarray}
where the sign $\pm$ depends on whether $\alpha^{(p)}$ has the orientation 
induced by $\beta^{(p+1)}$ or not. To complete the proof we show that this
is equal to the r.h.s. of (\ref{2.17}). Setting $\alpha^{(p)}=
 {\lbrack}v_0,\dots,v_p\rbrack$ we find
\begin{eqnarray*}
<\ast^Kd^K\alpha^{(p)}\,,\,\widehat{\beta^{(p+1)}}>=<d^K\alpha^{(p)}\,,\,
\beta^{(p+1)}>=\sum_{v_{p+1}}<{\lbrack}v_{p+1},v_o,\dots,v_p\rbrack\,,\,
\beta^{(p+1)}> \\
=(-1)^{p+1}\sum_{v_{p+1}}<{\lbrack}v_0,\dots,v_p,v_{p+1}\rbrack\,,\,
\beta^{(p+1)}>=(\ref{2.18})\,.\qquad\qquad\qed
\end{eqnarray*}

{\it Remark 2.8}.
The Whitney map generalises to a map between cohains and forms with 
coefficients in a flat vector bundle \cite{Muller}. The constructions and
results of this section generalise in an obvious way to that setting.

{\it Remark 2.9}.
Although we work exclusively with triangulations,
our techniques and results here and in the following
also go through for more general polyhedral 
decompositions $K$ of $M\,$, in particular for cubic 
decompositions. In these cases the barycentric subdivision and dual of the
decomposition of $M$ can be constructed in an analogous way to the simplicial
case. The resulting barycentric subdivision $BK$ will necessarily be a 
triangulation (even though $K$ is not), so the Whitney map $W^{BK}$ on 
which our constructions are based continues to be well-defined,
and it is straightforward to check that our constructions and results continue 
to hold.

\section{Linking numbers in a simplicial framework}

Recall the setup of the preceding section: $K$ is a simplicial complex
triangulating $M$ (${\dim}M=n$); $BK$ denotes its barycentric 
subdivision and $\widehat{K}$ denotes the dual triangulation. 
In this section the smoothness requirement on $M$ can be dropped, and all 
maps that we consider are continuous unless otherwise stated.
Let $N_1$ and $N_2$ be closed oriented manifolds 
with ${\dim}N_1\!=\!p\,{\le}\,n-1\,$,
${\dim}N_2=n-p-1\,$, triangulated by simplicial complexes $L_1$ and $L_2$
respectively. Let $f_K:L_1{\to}K$ be a simplicial map, i.e. a map 
$f_K:N_1{\to}M$ which maps each simplex of $L_1$ 
linearly onto a simplex of $K$. Let $g_{\widehat{K}}:\widehat{L_2}\to
\widehat{K}$ be a {\it dual-simplicial} map, i.e. a simplicial map
$g_{\widehat{K}}:BL_2{\to}BK$ which maps 
each cell in $\widehat{L_2}$ (considered as
a union of simplices in $BL_2$) onto a cell in $\widehat{K}$. (This determines
a map $g_{\widehat{K}}:N_2{\to}M$.)
In this section we show that in the simplicial framework
there is a natural expression $\mbox{lk}(f_K,g_{\widehat{K}})$ 
((\ref{3.1}) below) for the 
linking number of $f_K$ and $g_{\widehat{K}}$ when these maps are bounding 
(as defined below), and derive a formula for $\mbox{lk}(f_K,g_{\widehat{K}})$
in terms of the ingredients of the simplicial Hodge--de Rham theory 
developed in the last section (theorem 3.3).
We go on to study the linking number
$\mbox{lk}(f_K,g_K)$ of $f_K$ with a simplicial map $g_K:L_2{\to}K$ 
via the introduction of {\it simplicial framings}. A simplicial framing of
$g_K$ is essentially a homotopy $g$ from $g_K$ to a dual-simplicial map
$g_{\widehat{K}}:\widehat{L_2'}\to\widehat{K}\,$, $N_2{\to}M$ where $L_2'$
is another simplicial complex triangulating $N_2$. 
The point is that although a priori there is no natural expression 
for the linking number of $f_K$ and $g_K$ in the simplicial framework, 
we do obtain a natural expression for this linking number 
in terms of a simplicial
framing $g$ of $g_K\,$, namely $\mbox{lk}(f_K,g_{\widehat{K}})$
where $g_{\widehat{K}}$ is the dual-simplicial map homotopic to $g_K$ 
via $g\,$, provided that the images of $f_K$ and $g$ are disjoint in $M$.   
We prove that simplicial framings always exist in the 
main case of physical interest, namely when ${\dim}M=3$ and $N=S^1$
(theorem 3.6), and that in this case if 
$f_K$ and $g_K$ have disjoint images in $M$
then there is a simplicial framing $g$ of $g_K$ such that $f_K$ and 
$g$ have disjoint images in $M$. 
Intuitively one would expect $\mbox{lk}(f_K,g_{\widehat{K}})$ 
to be independent of the choice of simplicial
framing $g$ of $g_K$ with image disjoint from that of $f$ in $M$ (provided
that such a framing actually exists), thus providing a 
definition for $\mbox{lk}(f_K,g_K)$. We prove in proposition 3.9 that this
is in fact the case.

To illustrate and motivate the general expression (\ref{3.1}) below for the
linking number we first consider the case where ${\dim}M\!=\!3\,$, 
$N_1=N_2=S^1$ and $f_K$ and $g_{\widehat{K}}$ are embeddings of $S^1$ in $M$.
Then $g_{\widehat{K}}(S^1)$ is a union of 1-cells in $\widehat{K}\,$, 
each of which is the dual of a 2-simplex in $K$ as in the figure below:

$$
\epsfysize=5cm\epsfbox{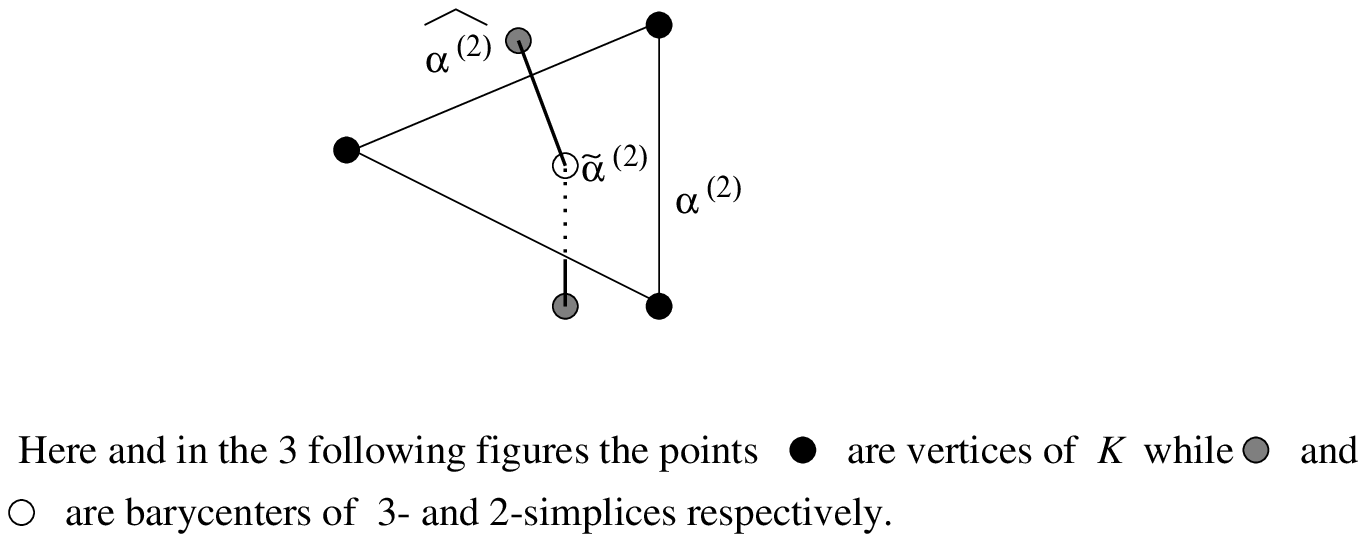}
$$

\noindent Thus if $D$ is a 2-dimensional surface in $M$ made up of a union
of 2-simplices of $K$ (considered as 2-cells in $M$) then 
$g_{\widehat{K}}(S^1)$ intersects $D$ transversely (if at all) at the 
barycenters of 2-simplices. Consider now the situation where $f_K(S^1)$
bounds such a surface $D$ as illustrated below:

$$
\epsfysize=4cm \epsfbox{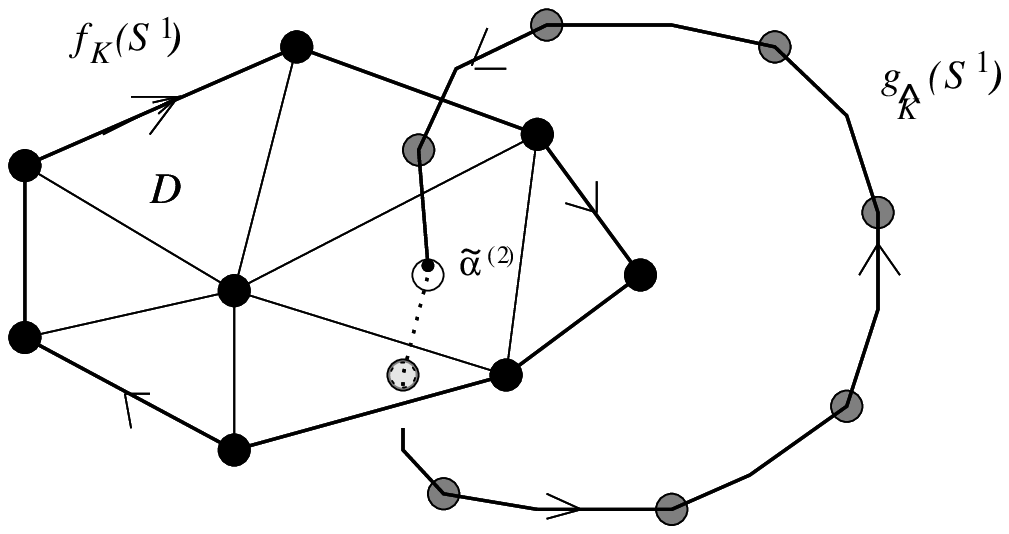}
$$

\noindent The linking number of $f_K(S^1)$ and $g_{\widehat{K}}(S^1)$ is 
now given by (see \cite[p.229]{BottTu}):
\begin{eqnarray} 
\mbox{lk}(f_K,g_{\widehat{K}})=\sum_{\widetilde{\alpha}^{(2)}\,\in\,
D{\cap}g_{\widehat{K}}(S^1)}\pm1
\label{3.05}
\end{eqnarray}
with sign determined as follows. The orientation of $f_K(S^1)$ induces an 
orientation for $D\,$, and thereby for
each 2-simplex $\alpha^{(2)}\in{}D\,$; this together with
the orientation of $M$ determines an orientation for $\widehat{\alpha^{(2)}}$.
If $\widetilde{\alpha}^{(2)}\in{}D\cap{}g_{\widehat{K}}(S^1)$ then the 
sign for the corresponding term in (\ref{3.05}) is positive if 
$\widehat{\alpha^{(2)}}$ has orientation compatible with 
$g_{\widehat{K}}(S^1)\,$, and negative otherwise. A succinct expression for
the linking number (\ref{3.05}) in the simplicial framework can now be 
obtained as follows. Let $\underline{D}\in{}C_2(K)$ denote the sum of the 
oriented 2-simplices making up $D\,$, then $\ast^K\underline{D}\in{}C_1
(\widehat{K})$ is the sum of the oriented 1-cells in $\widehat{K}$
(with orientation determined by $D$ and $M$) which intersect $D$ transversely
as shown in the figure below.

$$
\epsfbox{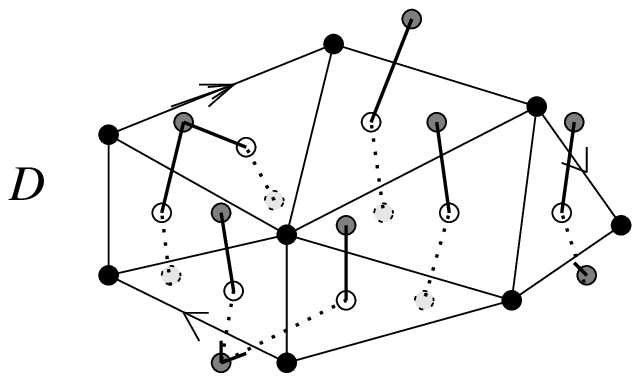}
$$

\noindent Let $\underline{g}_{\widehat{K}}\in{}C_1(\widehat{K})$ denote the
sum of the oriented 1-cells in $\widehat{K}$ which make up 
$g_{\widehat{K}}(S^1)\,$, then (\ref{3.05}) can be rewritten as
\begin{eqnarray}
\mbox{lk}(f_K,g_{\widehat{K}})=<\ast^K\underline{D}\,,
\underline{g}_{\widehat{K}}>
\label{3.07}
\end{eqnarray}

We now show how (\ref{3.07}) generalises for the general simplicial map
$f_K:L_1\to{}K\,$, $N_1\to{}M$ and dual-simplicial map
$g_{\widehat{K}}:\widehat{L_2}\to\widehat{K}\,$, $N_2\to{}M$ introduced at
the beginning of this section.
Let ${\lbrack}N_1\rbrack_{L_1}{\in}C_p(L_1)$ and 
${\lbrack}N_2\rbrack_{\widehat{L_2}}
{\in}C_p(\widehat{L_2})$ denote the orientation cycles of $N_1$ and $N_2$
determined by $L_1$ and $\widehat{L_2}$ respectively 
(i.e. the sum of all top degree
simplices (dual cells) with orientations induced by that of $N_1$ ($N_2$)).
The maps $f_K$ and $g_{\widehat{K}}$ induce chain maps $f_{K\#}:C_*(L_1){\to}
C_*(K)$ and $g_{\widehat{K}\#}:C_*(\widehat{L_2}){\to}C_*(\widehat{K})\,$,
defined by $f_{K\#}(\alpha^{(i)})\!=\!f_K(\alpha^{(i)})$ if $f_K(\alpha^{(i)})$
is of degree $i$ and $f_{K\#}(\alpha^{(i)})\!=\!0$ otherwise, and similarly
for $g_{\widehat{K}\#}$. 
For notational simplicity we set
$\underline{f}_K=f_{K\#}({\lbrack}N_1\rbrack_{L_1})\,\in\,C_p(K)$ and
$\underline{g}_{\widehat{K}}=g_{\widehat{K}\#}({\lbrack}N_2\rbrack_
{\widehat{L_2}})\,\in\,C_{n-p-1}(\widehat{K})$.

{\it Definition 3.1.} $f_K$ and $g_{\widehat{K}}$ above are
{\it bounding} if there are oriented manifolds $D_1$ and $D_2$ such that
for $i=1,2$ \hfill\break
(i) ${\partial}D_i=N_i$ and there is a simplicial complex triangulating
$D_i$ with $L_i$ as a subcomplex. \hfill\break
(ii) There are maps $h_i:D_i{\to}M$ with $h_1\Big|_{{\partial}D_1}
=f_K$ and $h_2\Big|_{{\partial}D_2}=g_{\widehat{K}}$.

{\it Lemma 3.2.}
Assume $f_K:L_1{\to}K$ above is bounding, with $h_1:D_1{\to}M$ as in 
definition 3.1. Then there is a simplicial complex $L_1'$ triangulating $D_1$
with $L_1$ as a subcomplex, and a simplicial map $h_1':L_1'{\to}K\,$,
$N_1{\to}M$ coinciding with $f_K$ on $L_1$.
Setting $\underline{h}_1':=h_{1\#}'({\lbrack}D_1\rbrack_{L_1'})
\in{}C_{p+1}(K)$ we have $\partial^K\underline{h}_1'=f_K$.

\noindent {\it Proof.} The generalised barycentric subdivision procedure 
described in \cite[\S16]{Munkres} provides $L_1'$ and $h_1'$ (a simplicial
approximation to $h_1$) with the required properties. It immediately follows
that $\partial^K\underline{h}_1'=h_{1\#}'(\partial^{L_1'}
{\lbrack}D_1\rbrack_{L_1'})=\underline{f}_K$. \qed

If $f_K$ is bounding and $h_1'$ is as in lemma 3.2 then the expression 
(\ref{3.07}) for the linking number generalises to
\begin{eqnarray}
\mbox{lk}(f_K,g_{\widehat{K}})=<\ast^K\underline{h}_1'\,,g_{\widehat{K}}>
\label{3.1}
\end{eqnarray}
This is clearly integer valued (since $\underline{h}_1'$ and 
$\underline{g}_{\widehat{K}}$ are chains over ${\bf Z}$) and can easily
be rewritten in an analogous way to (\ref{3.05}) (we omit the details).

A priori the expression (\ref{3.1}) depends on the choice of 
$\underline{h}_1'$ and does not require $g_{\widehat{K}}$ to be bounding.
However, the following theorem shows that $\mbox{lk}(f_K,g_{\widehat{K}})$
is independent of the choice of $h_1'$ when both $f_K$ and $g_{\widehat{K}}$
are bounding.

{\it Theorem 3.3.} 
If $f_K$ and $g_{\widehat{K}}$ above are bounding then
\begin{eqnarray}
\mbox{lk}(f_K,g_{\widehat{K}})=<\underline{f}_K\,,
(\ast^Kd^K)^{-1}\underline{g}_{\widehat{K}}>\,.
\label{3.2}
\end{eqnarray}

\noindent The proof requires the following

{\it Lemma 3.4.} 
Assume $g_{\widehat{K}}:\widehat{L_2}{\to}
\widehat{K}$ above is bounding, with $h_2:D_2{\to}M$ as in  definition 3.1.
Then there is a simplicial complex $L_2'$ triangulating $D_2$ with $BL_2$
as a subcomplex, and a simplicial map $h_2':L_2'{\to}BK\,$, $D_2{\to}M$
with the following properties: (i) $h_2'\Big|_{{\partial}D_2}=g_{\widehat{K}}$.
(ii) $h_2'$ maps the j-skeleton of $L_2'$ (considered as a subspace of $D_2$)
into the j-skeleton of $\widehat{K}$ (considered as a subspace of $M$)
$\forall\,j=0,1,\dots,n\!-\!p$. (iii) The chain 
$\underline{h}_2'\,\equiv\,h_{2\#}'({\lbrack}D_2
\rbrack_{L_2'}){\in}C_{n-p}(BK)$ can be considered as a chain in
$C_{n-p}(\widehat{K})\,$, and 
$\partial^{\widehat{K}}\underline{h}_2'=\underline{g}_{\widehat{K}}$.

\noindent {\it Proof.}
The cellular approximation theorem (see e.g.
\cite[\S2.3.2]{Fuks}) provides a map $h_2'':D_2{\to}M$ mapping
the j-skeleton of the given triangulation of $D_2$ into the j-skeleton of
$\widehat{K}$ $\,\forall\,j\!=\!0,1,\dots,n\!-\!p$ 
with $h_2''\Big|_{{\partial}D_2}\!=\!
g_{\widehat{K}}$ ($\,h_2''$ is a cellular approximation to $h_2$).
The generalised barycentric subdivision procedure \cite[\S16]{Munkres}
then provides a simplicial complex $L_2'$ triangulating $D_2$ with $BL_2$
as a subcomplex, and a simplicial map $h_2':L_2'{\to}BK$ (a simplicial
approximation to $h_2''$) coinciding with $g_{\widehat{K}}$ on $BL_2$.
Clearly $h_2'$ must also map the j-skeleton of $L_2'$ into the 
j-skeleton of $\widehat{K}$ $\,\forall\,j\!=\!0,1,\dots,n\!-\!p$. 
This shows (i) and (ii). 
Let $\widehat{\alpha^{(p)}}$ be an arbitrary $(n-p)$-dimensional oriented
dual cell in $\widehat{K}\,$, with $|\widehat{\alpha^{(p)}}|$ the corresponding
closed cell in $M$. 
Using (ii) and the fact that $h_2'$ is continuous it is quite
easy to see that the restricted map $h_2':\,(h_2')^{-1}
(|\widehat{\alpha^{(p)}}|)\to|\widehat{\alpha^{(p)}}|$ has a well-defined 
degree $\mbox{deg}(h_2';|\widehat{\alpha^{(p)}}|)$ when $(h_2')^{-1}
(|\widehat{\alpha^{(p)}}|){\subset}D_2$ is given the orientation induced by
$D_2$ and $|\widehat{\alpha^{(p)}}|{\subset}M$ the orientation induced by
$\alpha^{(p)}$. (We set $\mbox{deg}(h_2';|\widehat{\alpha^{(p)}}|)=0$
if $(h_2')^{-1}(|\widehat{\alpha^{(p)}}|)$ is empty.) 
Then for arbitrary $(n-p)$-simplex $\beta={\lbrack}\widetilde{\alpha}^{(p)},
\widetilde{\alpha}^{(p+1)},\dots,\widetilde{\alpha}^{(n)}\rbrack$ in
$BK$ contained in $|\widehat{\alpha^{(p)}}|$ with orientation compatible with
$\widehat{\alpha^{(p)}}$ we have $<\underline{h}_2'\,,
\beta>=\mbox{deg}(h_2';|\widehat{\alpha^{(p)}}|)$ independent of $\beta\,$,
so $\underline{h}_2'$ can indeed be considered as a
chain in $C_{n-p}(\widehat{K})$ (i.e. $\underline{h}_2'=
\sum_{\alpha^{(p)}{\in}K}\mbox{deg}(h_2';|\widehat{\alpha^{(p)}}|)
\widehat{\alpha^{(p)}}\,$). Since $\partial^{BK}h_{2\#}'({\lbrack}D_2
\rbrack_{L_2'})=h_{2\#}'(\partial^{L_2'}{\lbrack}D_2\rbrack_{L_2'})=
h_{2\#}'({\lbrack}N_2\rbrack_{BL_2})$ it is clear from (i) that
$\partial^{\widehat{K}}\underline{h}_2'=
\underline{g}_{\widehat{K}}$. \qed

\noindent {\it Proof of theorem 3.3.} By proposition 2.4 the r.h.s. of
(\ref{3.2}) is well-defined if
\begin{eqnarray}
<((\ast^Kd^K)^*)^{-1}\underline{f}_K\,,\underline{g}_{\widehat{K}}>=
(-1)^{np+1}<(\ast^{\widehat{K}}d^{\widehat{K}})^
{-1}\underline{f}_K\,,\underline{g}_{\widehat{K}}>
\label{3.3}
\end{eqnarray}
is well-defined. To see that (\ref{3.3}) is well-defined 
note that by lemma 3.2 and proposition 2.4
the equation $\ast^{\widehat{K}}d^{\widehat{K}}y=\underline{f}_K$ 
has a solution $y=(-1)^{np+1}\ast^K\underline{h}_1'\,$, and that if $y'$
is another solution then by lemma 3.4
\begin{eqnarray*}
<y-y'\,,\underline{g}_{\widehat{K}}>&=&
<y-y'\,,\partial^{\widehat{K}}\underline{h}_2'> \\
&=&<(\ast^{\widehat{K}})^{-1}(\ast^{\widehat{K}}d^{\widehat{K}}(y-y'))\,,
\underline{h}_2'>=0\,.
\end{eqnarray*}
Hence the r.h.s. of (\ref{3.2}) is well-defined, and can be written as
\begin{eqnarray*}
<\underline{f}_K\,,(\ast^Kd^K)^{-1}\underline{g}_{\widehat{K}}>
\stackrel{(\ref{3.3})}{=}(-1)^{np+1}<y\,,\underline{g}_{\widehat{K}}>
=<\ast^K\underline{h}_1'\,,\underline{g}_{\widehat{K}}> 
\stackrel{(\ref{3.1})}{=}\mbox{lk}(\underline{f}_K,\underline{g}_{\widehat{K}})
\quad\qed
\end{eqnarray*}

In the remainder of this section we study the linking number 
of $f_K$ with another simplicial map $g_K:L_2{\to}K\,$, $N_2{\to}M$.
In this case the expression (\ref{3.1}) is not well-defined.
(Also, $h_1'(D_1)$ and $g_K(N_2)$ need not intersect transversely
in $M$.) In order to obtain an expression for the linking number
within the simplicial framework in this case 
we introduce a particular class of framings:

{\it Definition 3.5.} Let $L$ be a simplicial complex triangulating
a manifold $N$ and let $\gamma_K:L{\to}K\,$, $N{\to}M$ be a simplicial map.
A {\it simplicial framing} of $\gamma_K$ is a homotopy $\gamma:N\times
\lbrack0,1\rbrack{\to}M$ from $\gamma_K$ to a dual-simplicial map
$\gamma_{\widehat{K}}:\widehat{L'}\to\widehat{K}\,$, $N{\to}M$ where $L'$
is another simplicial complex triangulating $N\,$, together with a 
simplicial complex $L''$ triangulating $N\times\lbrack0,1\rbrack$ with
$BL$ and $BL'$ as subcomplexes.

Intuitively, if $g:N_2\times\lbrack0,1\rbrack
{\to}M$ is a simplicial framing of $g_K$ 
and $f_K(N_1)$ and $g(N_2\times\lbrack0,1\rbrack)$
are disjoint in $M$ then the linking number of $f_K$ and $g_K$ is 
$\mbox{lk}(f_K,g_{\widehat{K}})$ (where $g_{\widehat{K}}:=
g\Big|_{N_2\times\{1\}}$). We must verify that this really
is independent of the choice of simplicial framing $g$
for $g_K$. But first, to show that this is relevant, we must show that
simplicial framings actually exist. We will show that they always exist
in the main case of physical interest, namely when ${\dim}M=3$ and $N=S^1$.
It is possible that they always exist in more general circumstances but at
present we have not proved this.

{\it Theorem 3.6.} If ${\dim}M=3$ and $L$ is a simplicial complex
triangulating $S^1$ then every simplicial map $\gamma_K:L{\to}K\,$, 
$S^1{\to}M$ has a simplicial framing.

\noindent {\it Proof.} We will show that a simplicial framing for $\gamma_K$
can be constructed after choosing a sequence of 3-simplices associated 
with $\gamma_K(S^1)$ as indicated in the figure below

$$
\epsfxsize=14cm \epsfbox{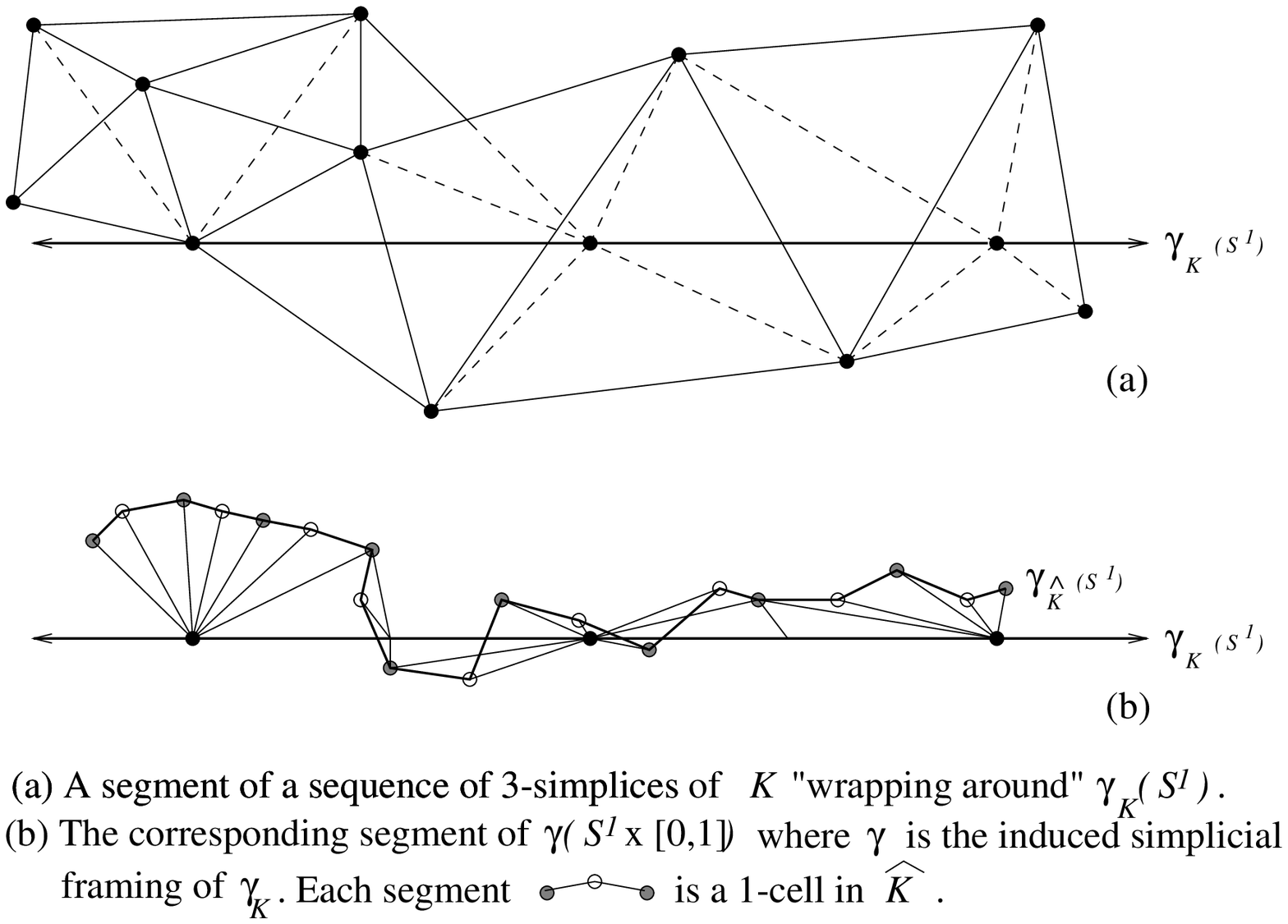}
$$

\noindent {\it Step 1.} We take $S^1$ to be $\lbrack0,1\rbrack$ with
endpoints identified so that the vertices of the triangulation $L$ are
$0\!=\!v_1<v_2<\cdots<v_N<v_{N+1}\!=\!1\,$ 
($v_1\!=\!v_{N+1}$ in $S^1$). Define $1{\le}i_1<i_2<\cdots<i_r<i_{r+1}{\le}\,
N\!+\!1$ by the condition that all $v_i$
with $i_j{\le}\,i<i_{j+1}$ are mapped by $\gamma_K$ to the same vertex $w_j$
in $K\,$, with $w_j{\ne}w_{j+1}$ $\forall\,j{\le}r$ and $w_{r+1}\!=\!w_1$.
Then ${\lbrack}w_j,w_{j+1}\rbrack$ is a 1-simplex in $K\,$ $\forall\,j\!=\!1,
\dots,r$ and $\gamma_K(S^1)$ is their union (considered as 1-cells in $M$).
Choose a 3-simplex $\alpha^{(3)}$ of $K$ in $\overline{\mbox{St}({\lbrack}
w_r,w_1\rbrack)}$. Choose a sequence of 3-simplices $\{\alpha_{1l}^{(3)}\}_{
l=1,\dots,l_1}$ in $\overline{\mbox{St}(w_1)}$ such that $\alpha_{11}^{(3)}=
\alpha^{(3)}$ and $\alpha_{1l_1}^{(3)}$ is contained in $\overline{\mbox{St}
({\lbrack}w_1,w_2\rbrack)}$ and $\alpha_{1l}^{(3)}\cap\alpha_{1(l+1)}^{(3)}$
is a 2-simplex $\alpha_{1l}^{(2)}\,$ $\forall\,l=1,\dots,l_1\!-\!1$.
(This is possible because otherwise $\overline{\mbox{St}(w_1)-\mbox{St}(
{\lbrack}w_r,w_1\rbrack)}$ would consist of two or more components meeting
only at points or line segments, in contradiction to the fact that 
$\overline{\mbox{St}(w_1)-\mbox{St}({\lbrack}w_r,w_1\rbrack)}$ is homeomorphic
to the closed ball $D^3$.) 
By discarding elements in the sequence if necessary, we
can assume that $\{\alpha_{1l}^{(3)}\}_{l=2,\dots,l_1-1}$ are all distinct and
contained in $\overline{\mbox{St}(w_1)-\{\mbox{St}({\lbrack}w_r,w_1\rbrack)
\cup\mbox{St}({\lbrack}w_1,w_2\rbrack)\}}$.
Proceeding inductively for $j\!=\!2,\dots,r$ 
we choose a sequence of 3-simplices
$\{\alpha_{jl}^{(3)}\}_{l=1,\dots,l_j}$ such that $\alpha_{j1}^{(3)}=
\alpha_{(j-1)l_{(j-1)}}^{(3)}\;$, $\,\alpha_{jl_j}^{(3)}\subset\overline{
\mbox{St}({\lbrack}w_j,w_{j+1}\rbrack)}\;$, 
$\,\{\alpha_{jl}^{(3)}\}_{l=2,\dots,l_j-1}$ are all distinct and contained in
$\overline{\mbox{St}(w_j)-\{\mbox{St}({\lbrack}w_{j-1},w_j\rbrack)
\cup\mbox{St}({\lbrack}w_j,w_{j+1}\rbrack)\}}$ and $\alpha_{jl}^{(3)}\cap
\alpha_{j(l+1)}^{(3)}$ is a 2-simplex $\alpha_{jl}^{(2)}$ 
$\,\forall\,j\!=\!1,\dots,l_j\!-\!1$.
Choose a sequence of distinct 3-simplices $\{\alpha_{(r+1)l}^{(3)}\}_{l=1,
\dots,l_{r+1}}$ in $\overline{\mbox{St}({\lbrack}w_r,w_1\rbrack)}$ such
that $\alpha_{(r+1)1}^{(3)}\!=\!\alpha_{rl_r}^{(3)}\;$, 
$\,\alpha_{(r+1)l_{r+1}}^{(3)}\!=\!\alpha_{11}^{(3)}$ 
and $\alpha_{(r+1)l}^{(3)}\cap\alpha_{(r+1)(l+1)}^{
(3)}$ is a 2-simplex $\alpha_{(r+1)l}^{(2)}$ 
$\,\forall\,l\!=\!1,\dots,l_{r+1}\!-\!1$.
(Again this is possible because otherwise we obtain a contradiction to 
the fact that $\overline{\mbox{St}({\lbrack}w_r,w_1\rbrack)}$ is homeomorphic
to $D^3$.)
Finally, choose a collection 
$\{x_{jl}\,|\,l\!=\!1,\dots,l_j\!-\!1\,;\,j\!=\!1,\dots,r\!+\!1\}$
of distinct points in ${\lbrack}0,1)$ so that 
$x_{11}\!=\!0\,$, $x_{jl}\!<\!x_{j'l'}$ for $j\!<\!j'$ and 
$x_{jl}\!<\!x_{jl'}$ for $l\!<\!l'$. Set $x_{jl_j}\!:=\!x_{(j+1)0}$
$\,\forall\,j\!=\!1,\dots,r\;$, 
$\,x_{(r+1)l_{r+1}}\!:=\!1$ and $y_{jl}\!:=\!\frac{1}{2}
(x_{jl}+x_{j(l+1)})$ $\,\forall\,l\!=\!1,\dots,l_j\!-\!1\;$ 
$\forall\,j\!=\!1,\dots,r\!+\!1$.
Taking $S^1$ to be $\lbrack0,1\rbrack$ with endpoints identified let $L'$
be the triangulation of $S^1$ with vertices 
$\{y_{jl}\,|\,l\!=\!1,\dots,l_j\!-\!1\,;\,j\!=\!1,\dots,r\!+\!1\}\,$; 
then $BL'$ has vertices $\{x_{jl},y_{jl}\}$ and $\widehat{L'}$
has vertices $\{x_{jl}\}$. A simplicial map $\gamma_{\widehat{K}}:BL'{\to}
BK$ is now defined by $\gamma_{\widehat{K}}(x_{jl}):=\widetilde{\alpha}_{jl}^{
(3)}\,$ and $\gamma_{\widehat{K}}(y_{jl}):=\widetilde{\alpha}_{jl}^{(2)}$.
This determines a dual-simplicial map 
$\gamma_{\widehat{K}}:\widehat{L'}\to\widehat{K}$
since the dual cell $\widehat{{\lbrack}y_{jl}\rbrack}={\lbrack}x_{jl},y_{jl}
\rbrack\cup{\lbrack}y_{jl},x_{j(l+1)}\rbrack$ in $\widehat{L'}$ is mapped by
$\gamma_{\widehat{K}}$ to the dual cell ${\lbrack}\widetilde{\alpha}_{jl}^{
(3)},\widetilde{\alpha}_{jl}^{(2)}\rbrack\cup\lbrack\widetilde{\alpha}_{jl}^{
(2)},\widetilde{\alpha}_{j(l+1)}^{(3)}\rbrack=\widehat{\alpha_{jl}^{(2)}}$
in $\widehat{K}$.

{\it Step 2.} Let $u_{i_j}'$ denote the centerpoint of ${\lbrack}v_{i_j},
v_{i_{j+1}}\rbrack$ $\forall\,j=1,\dots,r$ and define the triangulation
$L_{\frac{1}{2}}$ of 
$S^1\times\lbrack\frac{1}{2},1\rbrack$ as indicated in the figures below:

$$
\epsfxsize=16cm \epsfbox{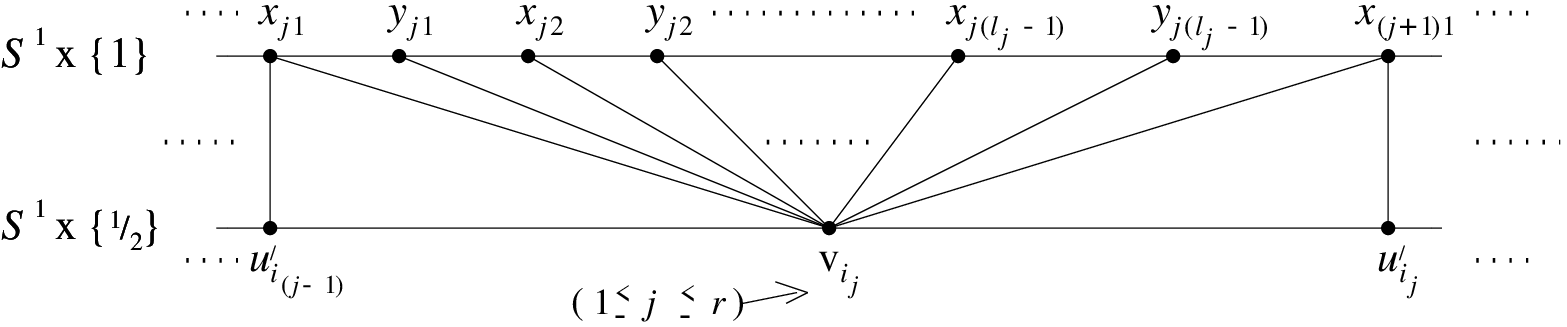}
$$

$$
\epsfxsize=16cm \epsfbox{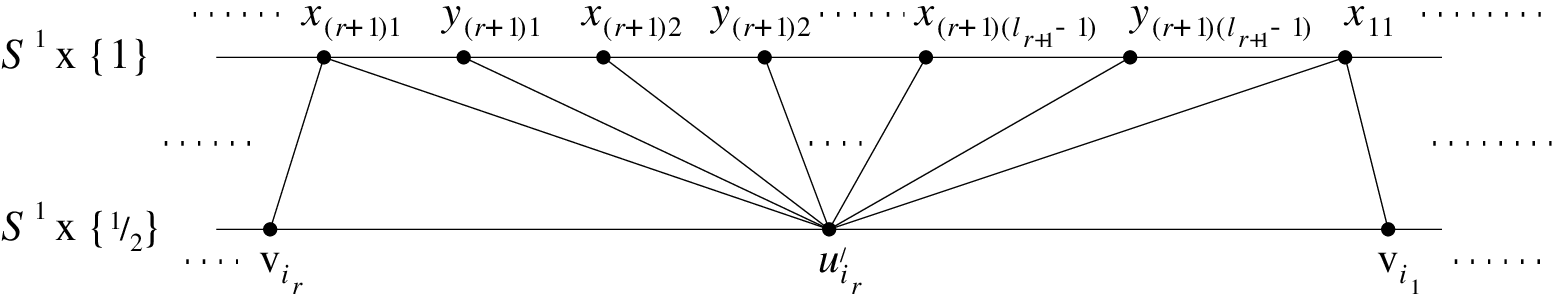}
$$

\noindent A simplicial map 
$\gamma_{\frac{1}{2}}:L_{\frac{1}{2}}{\to}BK$ is now defined by
$\gamma_{\frac{1}{2}}(x_{jl})\!:=\!\gamma_{\widehat{K}}(x_{jl})\;$, 
$\,\gamma_{\frac{1}{2}}(y_{jl})\!=\!\gamma_{\widehat{K}}(y_{jl})\;$, 
$\,\gamma_{\frac{1}{2}}(v_{i_j})\!=\!\gamma_K(v_{i_j})\!=\!w_j$ and 
$\gamma_{\frac{1}{2}}(u_{i_j}')\!=$the centerpoint of
${\lbrack}w_j,w_{j+1}\rbrack$. (It is easily checked that with the choices
above $\gamma_{\frac{1}{2}}$ maps 3-simplices of 
$L_{\frac{1}{2}}$ onto 3-simplices of
$BK$.) Let $u_i$ denote the centerpoint of ${\lbrack}v_i,v_{i+1}\rbrack$
$\forall\,i\!=\!1,\dots,N$ and define the the triangulation $L_0$ of 
$S^1\times\lbrack0,\frac{1}{2}\rbrack$ as indicated in the figure below
(with $v_{i_{r+1}}:=v_1$):

$$
\epsfxsize=12.5cm \epsfbox{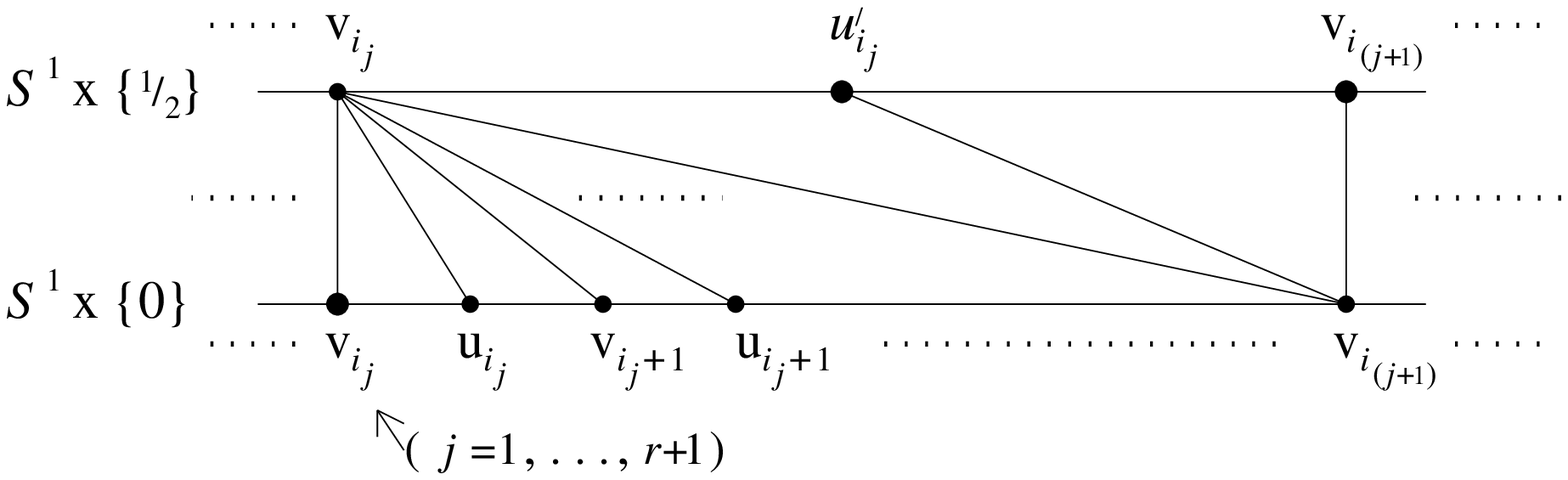}
$$

\noindent A simplicial map 
$\gamma_0:L_0{\to}BK$ is now defined by $\gamma_0(v_i)\!=\!
\gamma_K(v_i)\;$, $\gamma_0(u_{i_j}')\!=\!\gamma_{\frac{1}{2}}(u_{i_j})$ and
$\gamma_0(u_i)\!=$the centerpoint between $\gamma_K(v_i)$ and 
$\gamma_K(v_{i+1})$. Clearly $L_0$ and $L_{\frac{1}{2}}$ coincide on 
$S^1\times\{\frac{1}{2}\}\,$, as do $\gamma_0$ 
and $\gamma_{\frac{1}{2}}\,$, thus
determining a triangulation $L''=L_0{\ast}L_{\frac{1}{2}}$ of 
$S^1\times\lbrack0,1\rbrack$ and a simplicial map $\gamma:L''{\to}BK\,$,
and thereby a map $\gamma:S^1\times\lbrack0,1\rbrack{\to}M$.
It is clear from the constructions that $\gamma$ and $L''$ satisfy 
definition 3.5. \qed

Let $f_K:L_1{\to}K\,$, $N_1{\to}M$ and $g:L_2{\to}K\,$, $N_2{\to}M$ be
simplicial maps as above (${\dim}N_1\!=\!p\,$, ${\dim}N_2\!=\!n\!-\!p\!-\!1$).
Let $\mbox{St}_{BK}(f_K(N_1))$ denote the open region of $M$ consisting
of the union of all $\mbox{St}_{BK}(v)$ where $v$ is a vertex of $BK$ contained
in $f_K(N_1)\,$, and let $\mbox{St}_{BK}(g_K(N_2))$ be defined similarly.

{\it Definition 3.6.} A simplicial framing $g$ for $g_K$ is 
{\it strongly disjoint} from $f_K$ if the image of $g$ in $M$ is disjoint
from $\mbox{St}_{BK}(f_K(N_1))$.

{\it Observation 3.8.} 
(i) $f_K(N_1)\subseteq\mbox{St}_{BK}(f_K(N_1))$. 
(ii) $M-\mbox{St}_{BK}(f_K(N_1))$ has a cell decomposition given by 
$\widehat{K}$ with the cells $\{\widehat{v}\;|\;\mbox{$v$ vertex in $K$ 
contained in $f_K(N_1)$}\}$ removed. 
(iii) If $f_K(N_1){\cap}g_K(N_2)=\emptyset$ then $\mbox{St}_{BK}(f_K(N_1))
\cap\mbox{St}_{BK}(g_K(N_2))=\emptyset$. 
(iv) If ${\dim}M\!=\!3\;$, $N_1\!=\!N_2\!=\!S^1\;$, 
$f_K(S^1){\cap}g_K(S^1)=\emptyset$
and $g$ is a simplicial framing for $g_K$ constructed as in the proof
of theorem 3.6 then $g$ is strongly disjoint from $f_K$. 

{\it Proposition 3.9.}
Let $f_K$ and $g_K$ be simplicial maps as above, 
and assume that they are bounding with $f_K(N_1){\cap}g_K(N_2)=
\emptyset$. If there exists a simplicial framing $g$ for $g_K$ strongly
disjoint from $f_K$ (as is always the case when ${\dim}M\!=\!3$ and 
$N_1\!=\!N_2\!=\!S^1\,$, cf. theorem 3.6 and observation 3.8(iv)) then 
$\mbox{lk}(f_K,g_K)\!:=\!\mbox{lk}(f_K,g_{\widehat{K}})$ 
is independent of the choice of 
such a framing $g$. (Here $g_{\widehat{K}}$ is the dual-simplicial map
determined by $g$ as specified in definition 3.5. Note that $g_{\widehat{K}}$
is bounding since $g_K$ is, so $\mbox{lk}(f_K,g_{\widehat{K}})$ is 
well-defined.)

\noindent {\it Proof.} 
Set $M_f:=M-\mbox{St}_{BK}(f_K(N_1))$.
Let $\widetilde{g}$ be another simplicial framing for $g_K$ 
which is also strongly
disjoint from $f_K$. We can assume $\widetilde{g}$ is a map
$\widetilde{g}:N_2\times\lbrack-1,0\rbrack{\to}M_f$ with $g_K=
\widetilde{g}\Big|_{N_2\times\{0\}}$ and $\widetilde{g}_{\widehat{K}}=
\widetilde{g}\Big|_{N_2\times\{-1\}}:N_2{\to}M_f\,$, $\widehat{\widetilde
{L_2'}}{\to}\widehat{K}$ a dual-simplicial map for some simplicial complex
$\widetilde{L_2'}$ triangulating $N_2$. Let $L_2''$ and $\widetilde{L_2''}$
be the triangulations of $N\times\lbrack0,1\rbrack$ and
$N\times\lbrack-1,0\rbrack$ associated with $g$ and $\widetilde{g}$
respectively, as specified by definition 3.5. This data determines a 
dual-simplicial map $\widetilde{g}_{\widehat{K}}{\cup}g_{\widehat{K}}:
\widehat{\widetilde{L_2'}}\cup\widehat{L_2'}{\to}\widehat{K}\,$,
$N_2\times\{-1,1\}{\to}M_f$ which is bounding: The map $h_2:N_2\times
\lbrack-1,1\rbrack{\to}M_f$ required by definition 3.1 is defined by
$h_2\Big|_{N_2\times\lbrack-1,0\rbrack}=\widetilde{g}$ and 
$h_2\Big|_{N_2\times\lbrack0,1\rbrack}=g\,$, and the triangulation of
$D_2=N_2\times\lbrack-1,1\rbrack$ is $\widetilde{L_2''}{\ast}L_2''$.
Lemma 3.4 continues to hold in this case with $M$ replaced by $M_f$.
(The proof goes through unchanged due to observation 3.8(ii).)
Thus we have a triangulation $L_2'''$ (denoted $L_2'$ in lemma 3.4) of
$N_2\times\lbrack-1,1\rbrack$ and a simplicial map $h_2':L_2'''{\to}BK\,$,
$N_2\times\lbrack-1,1\rbrack{\to}M_f$ such that
$\underline{h}_2'{\in}C_{n-p}(BK)$ can be considered as a chain in 
$C_{n-p}(\widehat{K})$ and
\begin{eqnarray}
\partial^{\widehat{K}}\underline{h}_2'=
\underline{g}_{\widehat{K}}-\underline{\widetilde{g}}_{\widehat{K}}\,.
\label{3.4}
\end{eqnarray}
It follows from (\ref{3.4}), theorem 3.3 and proposition 2.4 that
(modulo an irrelevant sign)
\begin{eqnarray}
\mbox{lk}(f_K,g_{\widehat{K}})-\mbox{lk}(f_K,\widetilde{g}_{\widehat{K}})
&=&<\underline{f}_K\,,(\ast^Kd^K)^{-1}(\underline{g}_{\widehat{K}}-
\underline{\widetilde{g}}_{\widehat{K}})> \nonumber \\
&=&\pm<\underline{f}_K\,,(\partial^{\widehat{K}}
\ast^K)^{-1}\partial^{\widehat{K}}\underline{h}_2'> \nonumber \\
&=&\pm<\underline{f}_K\,,\ast^{\widehat{K}}\underline{h}_2'>
\label{3.5}
\end{eqnarray}
Since $h_2'(N_2\times\lbrack-1,1\rbrack)\,{\subseteq}\,M\!-\!\mbox{St}_{BK}
(f_K(N_1))$ the vertices of $BK$ contained in $h_2'(N_2\times\lbrack-1,1
\rbrack)$ are disjoint from the vertices of $BK$ contained in $f_K(N_1)$
so (\ref{3.5}) vanishes. \qed

Finally, if $M$ has odd dimension $n\!=\!2p\!+\!1$ and $N$ is a closed oriented
p-dimensional manifold triangulated by a simplicial complex $L$ then
each simplicial framing $f$  for a simplicial map $f_K:L{\to}M\,$, $N{\to}M$
has a {\it self--linking number} 
$\,\mbox{lk}(f):=\mbox{lk}(f_K,f_{\widehat{K}})$.

\section{Continuum abelian Chern--Simons gauge theory}

In this section we review the evaluation of the partition function
\cite{Schwarz} \cite{W(Jones)} \cite{AdSe(PLB)} and Wilson v.e.v.'s of linked 
loops \cite{Poly} \cite{W(Jones)} in the continuum abelian Chern--Simons 
theory on closed 3-manifolds (and its generalisation to arbitrary closed
odd-dimensional manifolds). Our treatment includes certain features involving
the moduli space of flat $U(1)$ connections (when $\pi_1(M)$ is non-trivial)
not contained in the original 
treatments\footnote{While we were finishing the writeup of this paper
a preprint by M.~Manoliu \cite{Manoliu} appeared 
in which abelian Chern--Simons
theory is constructed as a topological field theory 
(in the sense of Witten--Atiyah)
via geometric quantisation. This preprint also contains a comprehensive
treatment of the standard path--integral evaluation of the partition 
function, along similar lines to the present review.}.

{\it The Setup.}
Let $P$ be a flat principal $U(1)$ bundle 
over a closed oriented 3-manifold $M$. 
(All bundles, manifolds, maps etc. in this section are smooth.)
The connections on $P$ form an affine vectorspace ${\cal C}$ modelled
on the space of 1-forms with values in the adjoint bundle, which is 
necessarily trivial, so given $A_0\in{\cal C}$ the elements $A$ of ${\cal C}$
are $A=A_0+2{\pi}i\omega\,$, $\omega\in\Omega^1(M)$. The abelian 
Chern--Simons action functional on ${\cal C}$ is defined by
\begin{eqnarray}
i{\lambda}S(A_0+2{\pi}i\omega):=i{\lambda}\int_Md\omega\wedge\omega
\label{4.1}
\end{eqnarray}
for flat connection $A_0\,$, where $\lambda\in{\bf R_+}$ is the coupling
parameter. This is independent of the choice of flat connection $A_0$
by Stokes theorem, since if $A_0'$ is another flat connection then
$d(A_0'-A_0)=0$. The group ${\cal G}$ of gauge transformations can be 
identified with the maps $g:M{\to}U(1)\,$, and acts on ${\cal C}$
by $g{\cdot}A=A+gdg^{-1}$. Locally $g\in{\cal G}$ is of the form
$g=e^{2{\pi}if}$ for ${\bf R}$--valued function $f\,$, leading to 
$gdg^{-1}=-2{\pi}idf\,$, so $d(gdg^{-1})=0$ globally and it follows from 
Stokes theorem that the action functional (\ref{4.1}) is gauge--invariant.
However, if $\pi_1(M)$ is non--trivial and $g$ has non--zero winding number
around loops in $M$ then globally $f$ is only defined modulo ${\bf Z}\,$,
since the integral of $(-2{\pi}i)^{-1}gdg^{-1}$ around a cycle in $M$ equals
the winding number of $g$ around the cycle\footnote{This detail was omitted
in the original treatments, where the gauge transformations were restricted
to those of the form $A{\mapsto}A\!+\!2{\pi}idf$.}.
Thus $(-2{\pi}i)^{-1}gdg^{-1}$ represents an element of 
$H_{dR}^1(M)_{\bf Z}\,$, where we are using

{\it Definition 4.1.} $H_{dR}^1(M)_{\bf Z}$ is the subgroup of $H_{dR}^1(M)$
consisting of the elements whose integrals around the cycles of $M$  
are ${\bf Z}$--valued. (Here $H_{dR}^q(M)$ denotes the $q$'th de Rham 
cohomology group.)

\noindent In the following we will use the fact that the de Rham map gives an 
isomorphism $H_{dR}^1(M)_{\bf Z}\,\cong\,\mbox{Free}H^1(M,{\bf Z})\!=\!
H^1(M,{\bf Z})$. (The last equality is the standard fact that $H^1(M,{\bf Z})$
has no torsion; this is clear for example from (\ref{6.1}) in \S6.)

Pick a metric on $M\,$, let ${\cal H}^q(M)$ denote the space of harmonic
$q$-forms and let $\phi_q:{\cal H}^q(M)\stackrel{\cong}{\to}H_{dR}^q(M)$
denote the isomorphism induced by the projection map $\ker(d_q){\to}
H_{dR}^q(M)$. 
The homomorphism
${\cal G}{\to}H_{dR}^1(M)_{\bf Z}\,$, $g\mapsto(-2{\pi}i)^{-1}gdg^{-1}$
induces an isomorphism of the homotopy equivalence
classes of ${\cal G}$ onto $H_{dR}^1(M)_{\bf Z}$ \cite{AtiyahBott}.
This has the following consequences: \hfill\break
(i) ${\cal G}\,\cong\,\exp(2{\pi}i\Omega^0(M)){\times}H_{dR}^1(M)_{\bf Z}$.
\hfill\break
(ii) The orbit of ${\cal G}$ through $A\in{\cal C}$ is 
\begin{eqnarray}
{\cal G}{\cdot}A\ {\cong}\ A+2{\pi}i(d_0\Omega^0(M){\times}H_{dR}^1(M)_{\bf Z})
\label{4.2}
\end{eqnarray}
and ${\cal G}/\widetilde{U(1)}\to{\cal G}{\cdot}A\;$, $g{\mapsto}
g{\cdot}A$ is bijective, where $\widetilde{U(1)}$ denotes the gauge
transformations which are constant on the connected components of $M$.
\hfill\break
(iii) Using the map $\phi_1$ the orbit space of ${\cal C}=A_0+2{\pi}i
\Omega^1(M)$ can be identified with
\begin{eqnarray}
{\cal C}/{\cal G}\ \cong\ A_0+2{\pi}i\left(\frac{H_{dR}^1(M)}
{H_{dR}^1(M)_{\bf Z}}\right)\oplus\ker(d_1)^{\perp}
\label{4.3}
\end{eqnarray}
and the moduli space of the flat $U(1)$ connections 
${\cal F}=A_0+2{\pi}i\ker(d_1)$ can be identified with
\begin{eqnarray}
{\cal F}/{\cal G}\ \cong\ A_0+2{\pi}i\left(\frac{H_{dR}^1(M)}
{H_{dR}^1(M)_{\bf Z}}\right)\,.
\label{4.3.5}
\end{eqnarray}
Since $H_{dR}^1(M)\cong{\bf R}^{{\dim}H^1}$ and $H_{dR}^1(M)_{\bf Z}
\cong{\bf Z}^{{\dim}H^1}$ (where
${\dim}H^q\equiv{\dim}H_{dR}^q(M)$) 
we have 
\begin{eqnarray}
{\cal F}/{\cal G}\ \cong\ {\bf T}^{{\dim}H^1}
\label{4.3.7}
\end{eqnarray}
where ${\bf T}^N\equiv{\bf R}^N/{\bf Z}^N$ is the $N-$torus.
Thus ${\cal F}/{\cal G}$ is a closed manifold with finite volume
determined by a choice of volume element for $H_{dR}^1(M)$.

In the following, if $L:V{\to}W$ is a linear map between vectorspaces
with inner products we will use the 
notation $|\det{}'(L)|:=\det{}'(L^*L)^{1/2}$ where $\det{}'(L^*L)$ is the 
product of the non--zero eigenvalues of $L^*L$. If $L$ is an isomorphism
then $|\det(L)|=|\det{}'(L)|$ depends only on the volume elements on
$V$ and $W$.
The metric on $M$ induces
metrics on $\Omega^q(M)\,$, ${\cal H}^q(M)$ and ${\cal G}$.
We choose volume elements for $H_{dR}^q(M)$
for $q=0,\dots,3$ (then $|\det(\phi_q)|$ is defined), and equip ${\cal C}$
with the metric determined by requiring that the map $\Omega^1(M)\to
{\cal C}\;$, $\omega{\mapsto}A_0\!+\!2{\pi}i\omega$ be an isometry.

{\it The partition function.} 
This is the formal object
\begin{eqnarray}
Z(M,\lambda):=\frac{1}{V({\cal G})}\int_{\cal C}{\cal D\/}A\,
e^{-i{\lambda}S(A)}=\frac{1}{V({\cal G})}\int_{{\cal C}/{\cal G}}
{\cal D\/}{\lbrack}A\rbrack\,V({\lbrack}A\rbrack)\,e^{-i{\lambda}S
({\lbrack}A\rbrack)}
\label{4.4}
\end{eqnarray}
where $V({\cal G})$ and $V({\lbrack}A\rbrack)$ are the formal volumes of 
${\cal G}$ and ${\lbrack}A\rbrack={\cal G}{\cdot}A$ respectively.
It follows from (i) and (ii) above that formally
\begin{eqnarray}
V({\lbrack}A\rbrack)=V({\cal G})V(\widetilde{U(1)})^{-1}|\det{}'(d_0)|
=V({\cal G})|\det(\phi_0)||\det{}'(d_0)|
\label{4.5}
\end{eqnarray}
Writing the action functional (\ref{4.1}) as 
$i{\lambda}S(A_0\!+\!2{\pi}i\omega)=
i{\lambda}<{\ast}d_1\omega\,,\omega>$ and using (\ref{4.5}) and (iii) above
we find
\begin{eqnarray}
Z(M,\lambda)=V({\cal F}/{\cal G})|\det(\phi_0)||\det(\phi_1)|^{-1}
|\det{}'(d_0)|\det{}'\left(\frac{i\lambda}{\pi}{\ast}d_1\right)^{-1/2}
\label{4.6}
\end{eqnarray}
where $V({\cal F}/{\cal G})$ is determined by 
the volume element on $H_{dR}^1(M)$
via (\ref{4.3.5}), independent of the metric on $M$. Evaluating the 
determinants in (\ref{4.6}) via zeta--regularisation as in \cite{AdSe(PLB)}
and using Hodge duality gives
\begin{eqnarray}
Z(M,\lambda)&=&
e^{-\frac{i\pi}{4}\eta({\ast}d_1)}\left(\frac{\lambda}{\pi}\right)^
{-{\dim}H^0/2+{\dim}H^1/2}V({\cal F}/{\cal G})\,\tau_{RS}(M,d)^{1/2}
\left(\frac{a_0}{a_1}\right)^{1/2} \nonumber \\
& &\label{4.7}
\end{eqnarray}
where $\eta({\ast}d_1)$ is the analytic continuation to zero of the 
eta--function of ${\ast}d_1$ and
\begin{eqnarray}
\tau_{RS}(M,d)=\prod_{q=0}^3\Bigl(\,|\det(\phi_q)||\det{}'(d_q)|\Bigr)^{(-1)^q}
\label{4.8}
\end{eqnarray}
(with $|\det'(d_3)|{\equiv}1$) is the Ray--Singer torsion of $(M,d)\,$, 
depending on the choices of volume elements for the spaces $H_{dR}^q(M)$
but independent of the choice of metric on $M$ \cite{RS}, and where for
$q=0,1$ we have set
\begin{eqnarray}
a_q:=|\det(\phi_q)||\det(\phi_{3-q})|
\label{4.8.5}
\end{eqnarray}
The quantities $a_q$ are independent of the metric on $M$. To see this we
choose an inner product in $H_{dR}^{3-q}(M)$ which reproduces the 
volume element and define $E_q:H^q(M){\to}H^{3-q}(M)$ by
$<E_q(a),b>:=\int_Ma{\wedge}b\,$, then $|\det(E_q)|$ is obviously 
metric--independent. But on the other hand $E_q$ coincides with the map
\begin{eqnarray*}
E_q:H_{dR}^q(M)\stackrel{\phi_q^{-1}}{\to}{\cal H}^q(M)\stackrel{\ast}{\to}
{\cal H}^{3-q}(M)\stackrel{(\phi_{3-q}^{-1})^*}{\to}H^{3-q}(M)
\end{eqnarray*}
since
\begin{eqnarray*}
\int_Ma{\wedge}b=\int_M\phi_q^{-1}(a)\wedge\phi_{3-q}^{-1}(b)
=<(\phi_{3-q}^{-1})^*(\ast\phi_q^{-1}(a)),b>
\end{eqnarray*}
so we have $|\det(E_q)|^{-1}=|\det(\phi_q)||\det(\phi_{3-q})|=a_q$.
(This also shows that given a volume element for 
$H_{dR}^q(M)$ we can choose the volume element for $H_{dR}^{3-q}(M)$ so that
$a_q=1$. Thus with appropriate choices of volume elements for $H_{dR}^*(M)$
the factor $\frac{a_0}{a_1}$ drops out of (\ref{4.7}).)
Thus the metric--dependence of the final expression (\ref{4.7}) for
$Z(M,\lambda)$ enters only through $\eta({\ast}d_1)$ in the phase, so
$|Z(M,\lambda)|$ is a topological invariant.

{\it The Wilson vacuum expectation value (v.e.v.).} 
This is the formal object
\begin{eqnarray}
<W(\gamma^{(1)},\dots,\gamma^{(r)}\,;\,n_1,\dots,n_r)>_{\lambda}&:=&
\frac{\int_{\cal C}{\cal D\/}A\,\Big\lbrack\prod_{j=1}^r\Phi(A,\gamma^{(j)},
n_j)\Big\rbrack\,e^{-i{\lambda}S(A)}}
{\int_{\cal C}{\cal D\/}A\,e^{-i{\lambda}S(A)}} \nonumber \\
& &\label{4.9}
\end{eqnarray}
where the $\gamma^{(j)}:S^1{\to}M$ are maps which
are bounding (i.e. there are maps $h^{(j)}:D_j{\to}M$ restricting to
$\gamma^{(j)}$ on ${\partial}D_j=S^1$)
with images $\gamma^{(j)}(S^1)$ mutually disjoint in
$M\,$, and $\Phi(A,\gamma^{(j)},n_j)$ is the monodromy of $A$ around 
$\gamma^{(j)}$ in the representation of $U(1)$ on ${\bf C}$ corresponding
to $n_j\in{\bf Z}$. The evaluation of (\ref{4.9}) proceeds heuristically as
follows\footnote{A mathematically rigorous model for $<W>_{\lambda}$
leading to the same final expression in terms of linking numbers as
below has been constructed in ref. \cite{Albeverio}.}. 
For $l=1,\dots,r$ let $\eta^{(j)}$ be the distribution given by
$\int_{S^1}\gamma^{(j)\ast}\omega=\int_M\omega\wedge\eta^{(j)}$ $\,\forall
\omega\in\Omega^1(M)$. We will treat $\eta^{(j)}$ heuristically as a 
2-form on $M$ vanishing away from $\gamma^{(j)}(S^1)\,$, then 
$\eta^{(j)}=d\omega^{(j)}$ where $\omega^{(j)}$ is the 1-form distribution
given by $\int_{D_j}h^{(j)\ast}\tau=\int_M\tau\wedge\omega^{(j)}$
$\,\forall\tau\in\Omega^2(M)$.
Writing $i{\lambda}S(A_0\!+\!2{\pi}i\omega)=
i{\lambda}<{\ast}d_1\omega\,,\omega>$ and
\begin{eqnarray*}
\Phi(A_0\!+\!2{\pi}i\omega,\gamma^{(j)},n_j)
&=&\Phi(A_0,\gamma^{(j)},n_j)\exp(2{\pi}in_j\int_{S^1}\gamma^{(j)\ast}
\omega) \\
&=&\Phi(A_0,\gamma^{(j)},n_j)\exp(2{\pi}in_j<\ast\eta^{(j)},\omega>)
\end{eqnarray*}
the formal evaluation of (\ref{4.9}) gives
\begin{eqnarray}
\lefteqn{<W(\gamma^{(1)},\dots,\gamma^{(r)};n_1,\dots,n_r)>_{\lambda}} 
\nonumber \\
&=&\left(\frac{1}{V({\cal F}/{\cal G})}\int_{{\cal F}/{\cal G}}
{\cal D\/}{\lbrack}A\rbrack\prod_{l=1}^r\Phi({\lbrack}A\rbrack,\gamma^{(l)},
n_l)\right)\exp\left(\frac{i\pi^2}{\lambda}\sum_{j\,m}n_jn_m<\ast\eta^{(j)},
({\ast}d_1)^{(-1)}\ast\eta^{(m)}>\right) \nonumber \\
&=&\exp\left(\frac{i\pi^2}{\lambda}\sum_{j\,m}n_jn_m\mbox{lk}(\gamma^{(j)},
\gamma^{(m)})\right)
\label{4.10}
\end{eqnarray}
where we have used the facts that
\begin{eqnarray*}
<\ast\eta^{(j)},({\ast}d_1)^{-1}\ast\eta^{(m)}>\,=\int_M\omega^{(m)}\wedge
\eta^{(j)}=\mbox{lk}(\gamma^{(j)},\gamma^{(m)})
\end{eqnarray*}
is the linking number of $\gamma^{(j)}$ and $\gamma^{(m)}$ (see p.230 of
\cite{BottTu}), and 
$\Phi(A,\gamma^{(l)},n_l)\!=\!1$ for flat connection
$A$ since the $\gamma^{(l)}$'s are
bounding and thereby trivial in $\pi_1(M)$. 
(In \S6 we will consider a situation where the integrand in the integral over
${\cal F}/{\cal G}$ in (\ref{4.10}) is a non-trivial constant.)
Finally, to obtain a well-defined
expression for $<W>_{\lambda}$ the self-linking numbers 
$\mbox{lk}(\gamma^{(j)},
\gamma^{(j)})$ in (\ref{4.10}) must be regularised ``by hand'' by introducing
a framing of each $\gamma^{(j)}$ \cite{Poly} \cite{W(Jones)}. Note that
the final expression (\ref{4.10}) is trivial when 
$\lambda\!=\!\frac{{\pi}p}{2}\,$,
$p\in{\bf Z}$. In \S6 we will see how $<W>_{\lambda}$ is non-trivial for
these discrete values of $\lambda$ when the $\gamma^{(j)}$'s represent
torsion elements in the homology of $M$ over ${\bf Z}$. 
(In the present case the $\gamma^{(j)}$'s represent the trivial
element in the homology of $M$ since they are bounding.)

{\it Framings and abelian BF theory.} 
The regularisation of self-linking numbers via framings of the 
$\gamma^{(j)}$'s can be incorporated into the theory from the beginning
(as opposed to putting it in by hand at the end) if we work with the 
doubled, ``twisted'' version of abelian Chern--Simons theory for two
independent connections $A\!=\!A_0\!+\!2{\pi}i\omega$ and 
$A'\!=\!A_0\!+\!2{\pi}i\omega'$
in ${\cal C}$ obtained by making a ``twist'' in the doubled action
\begin{eqnarray*}
i{\lambda}S(A)+i{\lambda}S(A')=
i\lambda\int_Md_1\omega\wedge\omega+d_1\omega'\wedge\omega'=
i\lambda\Big<\left(\begin{array}{c}\omega \\ \omega'\end{array}\right)\,,
\left(\begin{array}{cc}{\ast}d_1 & 0 \\ 0 & {\ast}d_1\end{array}\right)
\left(\begin{array}{c}\omega \\ \omega'\end{array}\right)\Big> \nonumber \\
\end{eqnarray*}
to obtain
\begin{eqnarray}
i\lambda\widetilde{S}(A,A'):=i\lambda
\Big<\left(\begin{array}{c}\omega \\ \omega'\end{array}\right)\,,
\left(\begin{array}{cc}0 & {\ast}d_1\\ {\ast}d_1 & 0\end{array}\right)
\left(\begin{array}{c}\omega \\ \omega'\end{array}\right)\Big>
=2i\lambda\int_Md\omega\wedge\omega'\,.
\label{4.12}
\end{eqnarray}
This is the action functional for the so-called abelian BF gauge theory.
It is easy to see that the partition function $\widetilde{Z}(M,\lambda)$
of this theory is given by
\begin{eqnarray}
\widetilde{Z}(M,\lambda)=|Z(M,\lambda)|^2
\label{4.13}
\end{eqnarray}
which is now completely independent of the choice of metric\footnote{The
absence of a complex phase factor in $\widetilde{Z}(M,\lambda)$ is because
the operator in the quadratic form (\ref{4.12}) has symmetric spectrum,
and therefore vanishing eta--function.}.
Let the $\gamma^{(j)}$'s now be framed loops, i.e. smooth maps
$\gamma^{(j)}:S^1\times\lbrack0,1\rbrack{\to}M\,$, with $\gamma_0^{(j)}:=
\gamma^{(j)}\Big|_{S^1\times\{0\}}:S^1{\to}M$ and
$\gamma_1^{(j)}:=\gamma^{(j)}\Big|_{S^1\times\{1\}}:S^1{\to}M$
such that $\gamma_0^{(j)}$ (and thereby also $\gamma_1^{(j)}$) is bounding
and the images $\gamma_0^{(j)}(S^1)\,$, $\gamma_1^{(m)}(S^1)$ are mutually
disjoint in $M$. The Wilson v.e.v. $<\widetilde{W}>_{\lambda}$ of the 
{\it framed} loops is now defined in a natural way in this theory 
by replacing $i{\lambda}S(A)$ and $\Phi(A,\gamma^{(j)},n_j)$ in (\ref{4.9}) by 
$i\lambda\widetilde{S}(A,A')$ and
$\Phi(A,\gamma_0^{(j)},n_j)\Phi(A',\gamma_1^{(j)},n_j)$ respectively,
and integrating (formally) over ${\cal C}\times{\cal C}$. 
Then the formal 
evaluation of $<\widetilde{W}>_{\lambda}$ leads to (\ref{4.10}) with
$\mbox{lk}(\gamma^{(j)},\gamma^{(m)})$ replaced by
$\mbox{lk}(\gamma_0^{(j)},\gamma_1^{(m)})+
\mbox{lk}(\gamma_1^{(j)},\gamma_0^{(m)})$.
Since $\mbox{lk}(\gamma_0^{(j)},\gamma_1^{(m)})=
\mbox{lk}(\gamma_0^{(j)},\gamma_0^{(m)})$
for $j{\ne}m$ and $\mbox{lk}(\gamma_0^{(s)},\gamma_1^{(s)})=
\mbox{lk}(\gamma^{(s)})$
is the self-linking number of $\gamma^{(s)}$ we obtain
\begin{eqnarray}
\lefteqn{<\widetilde{W}(\gamma^{(1)},\dots,\gamma^{(r)};n_1,\dots,n_r)>_
{\lambda}} \nonumber \\
&=&\exp\left(\frac{2i\pi^2}{\lambda}\left(\sum_{j{\ne}m}n_jn_mlk(
\gamma_0^{(j)},\gamma_0^{(m)})+\sum_{s=1}^rn_s^2lk(\gamma^{(s)})\right)\right)
\label{4.14}
\end{eqnarray}

Along with the feature described above, the abelian BF theory has another
advantage over the abelian Chern--Simons theory when it comes to constructing
simplicial versions of the theories: The simplicial analogue of the Hodge
star operator (i.e. the duality operator) can be incorporated in a natural
way into a simplicial version of the abelian BF theory since it has two 
independent coupled gauge fields. (We will see this explicitly in the next 
section.) This feature is essential if the same evaluation of the partition
function in a simplicial version of the theory is to lead to the same
final expression with R--torsion in place of Ray--Singer torsion.
Therefore we will not attempt to discretise the abelian Chern--Simons theory
directly, but instead construct a natural simplicial version of the 
abelian BF theory --this is done in the next section.

The abelian Chern--Simons theory and the corresponding abelian BF theory
generalise to abelian gauge theories on closed oriented manifolds $M$ of
arbitrary odd dimension with a flat vector bundle $E_{\rho}$ over $M$ 
determined by an orthogonal representation $\rho:\pi_1(M){\to}O(N)$.
The partition functions of the generalised theories can be formally evaluated 
by the technique of A.~Schwarz, leading to the square root of the 
Ray--Singer torsion $\tau_{RS}(M,d^{\rho})$ of the twisted de Rham complex
$\{\Omega^*(M,E_{\rho})\,,\,d^{\rho}\}$ \cite{Schwarz}\footnote{Strictly
speaking, the generalised theories are only generalisations of abelian 
Chern--Simons-- and BF--theories when the group ${\cal G}$ of gauge
transformations is replaced in the preceding by its identity-connected
component ${\cal G}_0\,$, since the field(s) in the generalised theories
transform under a gauge transformation by $\omega\to\omega+d^{\rho}\tau$.}.
Also, when the representation $\rho$ is trivial the Wilson v.e.v.'s of  
loops generalise and can be evaluated to obtain an expression in terms of
linking numbers as above.

\section{Simplicial abelian gauge theory}

In this section we present a canonical simplicial version of the abelian
BF theory with action (\ref{4.12}). Using the results of \S2--3
we show that the formal evaluation of the partition function and 
Wilson v.e.v.'s of framed loops (for simplicial framings of edge loops in the 
triangulation $K$) leads to expressions in terms of R--torsion and linking
numbers respectively, reproducing the continuum expressions\footnote{The
continuum expression for the partition function is reproduced when the 
value of the coupling parameter is 1; for values $\ne1$ a 
triangulation-dependent renormalisation of the coupling parameter is required
to reproduce this expression. 
The continuum expressions for the Wilson v.e.v.'s are reproduced for all
values of the coupling parameter (without the need for renormalisation).}.

{\it Simplicial action functional.} $A_0$ continues to denote the flat
connection in ${\cal C}$ chosen in the last section (the expressions below
for the partition function and Wilson v.e.v.'s are independent of the choice
of $A_0$). Define the embeddings
\begin{eqnarray}
\Psi_K:C^1(K)\hookrightarrow{\cal C}\ & &\ x{\mapsto}A_0+2{\pi}iW^{BK}(Bx)
\label{5.0.1} \\
\Psi_{\widehat{K}}:C^1(\widehat{K})\hookrightarrow{\cal C}\ & &\ y{\mapsto}
A_0+2{\pi}iW^{BK}(By) \label{5.0.2}
\end{eqnarray}
where the maps $B$ are given by (\ref{2.9})--(\ref{2.10}).
We construct the simplicial version of the abelian BF theory with action
(\ref{4.12}) by replacing the space of connections ${\cal C}\times{\cal C}$
by the subspace $\Psi_K(C^1(K))\times\Psi_{\widehat{K}}(C^1(\widehat{K}))$.
This is reminiscent of what happens in gauge fixing\footnote{I thank 
Prof. A.P.~Balachandran for this perspective.}.
The simplicial action functional is then $i\lambda\widetilde{S}_K\,$, where
$\widetilde{S}_K$ is the bilinear functional on $C^1(K){\times}
C^1(\widehat{K})$ given by
\begin{eqnarray}
\widetilde{S}_K(x,y):=\widetilde{S}(\Psi_K(x),\Psi_{\widehat{K}}(y))=
2\int_MdW^{BK}(Bx){\wedge}W^{BK}(By)
\label{5.1}
\end{eqnarray}
By theorem 2.6 we have
\begin{eqnarray}
\widetilde{S}_K(x,y)&=&\frac{1}{6}<\ast^Kd^Kx\,,y>=\frac{1}{6}<x\,,
\ast^{\widehat{K}}d^{\widehat{K}}y> \nonumber \\
&=&\frac{1}{12}\Big<\left(\begin{array}{c}x \\ y\end{array}\right)\,,
\left(\begin{array}{cc}0 & {\ast}^{\widehat{K}}
d^{\widehat{K}}\\ {\ast}^Kd^K & 0\end{array}\right)
\left(\begin{array}{c}x \\ y\end{array}\right)\Big> 
\label{5.2}
\end{eqnarray}
To avoid the numerical factor in (\ref{5.2}) we will take 
$\lambda':=12\lambda$ to be our coupling parameter in the simplicial theory.

{\it Simplicial gauge transformations.}
We define a group ${\cal G}_K\times{\cal G}_{\widehat{K}}$ acting on
$C^1(K){\times}$$C^1(\widehat{K})$ and leaving the simplicial action 
$\widetilde{S}_K$ invariant as follows. Let $\lbrack{\cal G}\rbrack$
denote the collection of homotopy equivalence classes of ${\cal G}$
and choose a representative $h_{{\lbrack}g\rbrack}\in{\lbrack}g\rbrack$
for each ${\lbrack}g\rbrack\in\lbrack{\cal G}\rbrack$. 
Then ${\cal H}_{\lbrack{\cal G}\rbrack}:=
\{h_{{\lbrack}g\rbrack}\}_{{\lbrack}g\rbrack\in\lbrack{\cal G}\rbrack}$
is a group with multiplication defined by $h_{{\lbrack}g_1\rbrack}{\cdot}
h_{{\lbrack}g_2\rbrack}:=h_{{\lbrack}g_1g_2\rbrack}$ (this will differ
in general from the usual multiplication of $h_{{\lbrack}g_1\rbrack}$
and $h_{{\lbrack}g_2\rbrack}$). Set $({\cal G}_K)_0:=\{e^{2{\pi}iW^K(u)}
\,|\,u{\in}C^0(K)\}\,$; with the usual multiplication this is an abelian
Lie group with Lie algebra $2{\pi}iW^K(C^0(K))$. We define ${\cal G}_K$ 
to be the product ${\cal G}_K:=({\cal G}_K)_0\cdot{\cal H}_{\lbrack{\cal G}
\rbrack}$ where $e^{2{\pi}iW^K(u)}{\cdot}h_{{\lbrack}g\rbrack}$ is defined
by the usual multiplication. ${\cal G}_K$ is a subset of ${\cal G}\,$,
although in general the embedding ${\cal G}_K\hookrightarrow{\cal G}$ is 
only a homomorphism modulo homotopy equivalence. We define the action of
$h\in{\cal G}_K$ on $x{\in}C^1(K)$ by
\begin{eqnarray}
h{\cdot}x=x+A_{dR}^K(\,(-2{\pi}i)^{-1}hdh^{-1})
\label{5.3}
\end{eqnarray}
where $A_{dR}^K:\Omega^*(M){\to}C^*(K)$ is the de Rham map. Since 
$d^K(A^K(hdh^{-1}))\!=\!A^K(d(hdh^{-1}))$$=0\,$ 
the simplicial action functional
is invariant under ${\cal G}_K$ due to (\ref{5.2}). 
Using (\ref{2.4}) we get 
\begin{eqnarray}
\left(e^{2{\pi}iW^K(u)}h_{{\lbrack}g\rbrack}\right){\cdot}x=
h_{{\lbrack}g\rbrack}{\cdot}x+d_0^Ku
\label{5.4}
\end{eqnarray}
Combining this with de Rham's theorem that $A_{dR}^K$ induces isomorphisms
$H_{dR}^1(M)\stackrel{\cong}{\to}H^1(K,{\bf R})\,$, 
$H_{dR}^1(M)_{\bf Z}\stackrel{\cong}{\to}H^1(K,{\bf Z})$
we see that the properties (i), (ii) and (iii) of ${\cal G}$ described in the
preceding section have analogues for ${\cal G}_K$. 
Set ${\cal H}^q(K):=\{x{\in}C^q(K)\,|\,d^Kx=(d^K)^*x=0\}$ (i.e. the simplicial
analogues of the harmonic forms), equipped with the canonical inner product
from $C^q(K)\,$, and let $\phi_q^K:{\cal H}^q(K)\stackrel{\cong}{\to}
H^q(K,{\bf R})$ denote the isomorphism induced by the projections
$\ker(d_q^K){\to}H^q(K,{\bf R})$. Equip $H^q(K,{\bf R})$ with the volume
element induced by that of $H_{dR}^q(M)$ via the de Rham map; 
then $|\det(\phi_q^K)|$ is defined. Define ${\cal H}^q(\widehat{K})\,$,
$\phi_q^{\widehat{K}}$ and $|\det(\phi_q^{\widehat{K}})|$ analogously.
${\cal G}_K$ has the following properties: \hfill\break
(i)$'$ ${\cal G}_K\ \cong\ \{e^{2{\pi}iW^K(u)}\,|\,u{\in}C^0(K)\}{\times}
H^1(K,{\bf Z})$. \hfill\break
(ii)$'$ The orbit ${\lbrack}x\rbrack$ of ${\cal G}_K$ through 
$x{\in}C^1(K)$ can be identified with
\begin{eqnarray}
{\cal G}_K{\cdot}x\ \cong\ x+(d_0^KC^0(K){\times}H^1(K,{\bf Z}))
\label{5.5}
\end{eqnarray}
and the map ${\cal G}_K/\widetilde{U(1)}\to{\cal G}_K{\cdot}x\;$, 
$h{\mapsto}h{\cdot}x$ is bijective, where $\widetilde{U(1)}_K:=
\{e^{2{\pi}iW^K(u)}\,|\,u\in\ker(d_0^K)\}$. \hfill\break
(iii)$'$ Using the map $\phi_1^K$ the orbit space of $C^1(K)$
can be identified with
\begin{eqnarray}
C^1(K)/{\cal G}_K\ \cong\ \left(\frac{H^1(K,{\bf R})}{
H^1(K,{\bf Z})}\right)\oplus\ker(d_1^K)^{\perp}
\label{5.6}
\end{eqnarray}
and the orbit space of ${\cal F}_K:=\ker(d_1^K)$ can be identified with
\begin{eqnarray}
{\cal F}_K/{\cal G}_K\ \cong\ \frac{H^1(K,{\bf R})}{H^1(K,{\bf Z})}
\label{5.6.5}
\end{eqnarray}

The group ${\cal G}_{\widehat{K}}$ cannot immediately be defined by analogy
with ${\cal G}_K$ because we do not have a Whitney map from $C^*(\widehat{K})$
to $\Omega^*(M)$. To get around this we construct a triangulation 
$\widehat{K}'$ of $M$ such that the vertices and 1-cells of $\widehat{K}$
are subsets of the vertices and 1-simplices of $\widehat{K}'\,$, and
$d_0^{\widehat{K}}$ coincides with the map
\begin{eqnarray}
C^0(\widehat{K}){\hookrightarrow}C^0(\widehat{K}')\stackrel{d_0^{\widehat{K}'}}
{\to}C^1(\widehat{K}'){\to}C^1(\widehat{K})
\label{5.7}
\end{eqnarray}
where the last map is the restriction. Then ${\cal G}_{\widehat{K}}$ is defined
by replacing $C^q(K)\,$, $W^K\,$, $d_0^K$ by $C^q(\widehat{K})\,$,
$W^{\widehat{K}'}\,$, $d_0^{\widehat{K}}$ respectively in the preceding
construction of ${\cal G}_K$. The group ${\cal G}_{\widehat{K}}$ and its action
on $C^1(\widehat{K})$ (defined by analogy with (\ref{5.3})) then has properties
completely analogous to the properties 
(\ref{5.4}), (i)$'$, (ii)$'$, (iii)$'$ of
${\cal G}_K\,$, and by (\ref{5.2}) the simplicial action functional is 
invariant under the action of ${\cal G}_{\widehat{K}}$ on $C^1(\widehat{K})$.
Briefly, $\widehat{K}'$ is constructed from the barycentric subdivision $BK$
as follows. Each $q$-simplex $\tau^{(q)}$ of $BK$ contained in a $3$-simplex
$\sigma^{(3)}$ of $K$ with a $(q-1)$-face $\tau^{(q-1)}$ of $\tau^{(q)}$
contained in a $2$-face $\sigma^{(2)}$ of $\sigma^{(3)}$ has a ``mirror
image'' $\tau^{(q)}{}'$ contained in the $3$-simplex $\sigma^{(3)}{}'$ of $K$
which shares $\sigma^{(2)}$ as a $2$-face. The union $\tau^{(q)}\cup
\tau^{(q)}{}'$ can be considered as an embedding of the standard $q$-simplex
into $M$ in a canonical way, as illustrated in the figure below
(after a planar projection):

$$
\epsfxsize=14cm \epsfbox{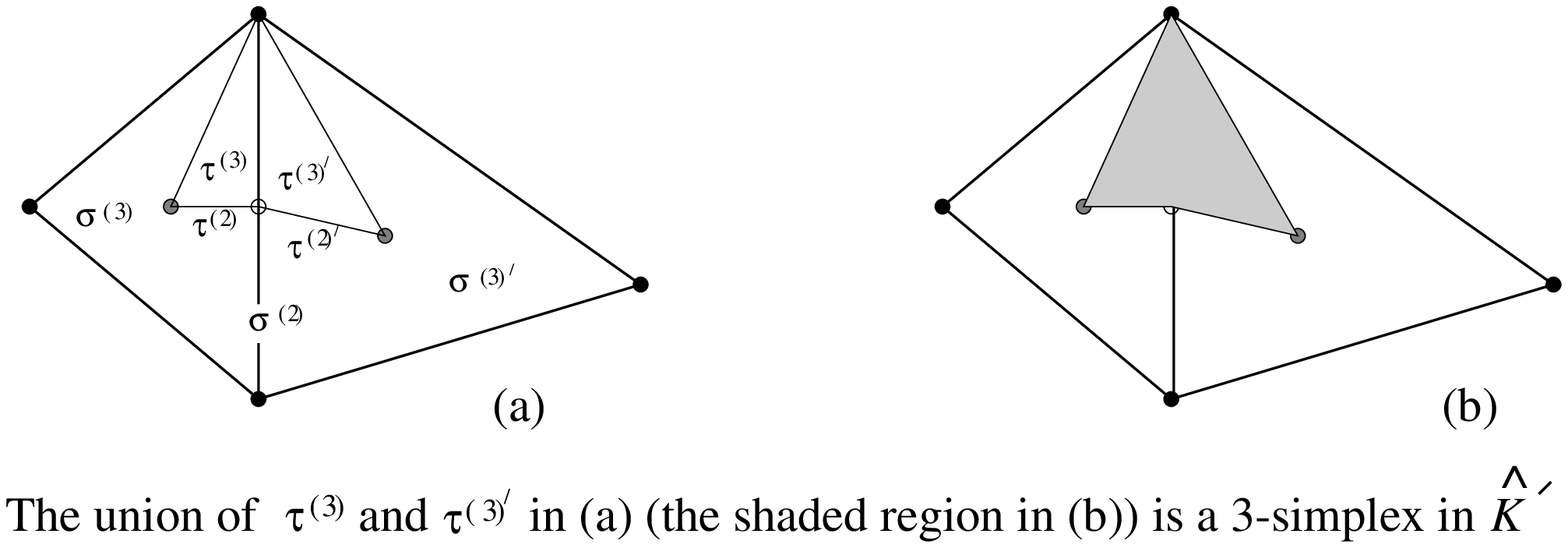}
$$

\noindent The vertices of $\widehat{K}'$ are the barycenters
of all the simplices in $K$ except the $2$-simplices. For $1{\le}q{\le}3$
the $q$-simplices of $\widehat{K}'$ (considered as cells in $M$) are the 
unions $\tau^{(q)}\cup\tau^{(q)}{}'$ of the $q$-simplices $\tau^{(q)}\,$,
$\tau^{(q)}{}'$ in $BK$ of the form described above, together with the
remaining $q$-simplices of $BK$. (For $q=3$ there are no remaining 
$3$-simplices.) Note that the $1$-cells of $\widehat{K}$ arise as the
unions $\tau^{(1)}\cup\tau^{(1)}{}'$. It is straightforward to check that
$\widehat{K}'$ thus defined is indeed a simplicial complex triangulating $M$
(smoothly on the complement of a set of measure zero) and that (\ref{5.7})
coincides with $d_0^{\widehat{K}}$ as required. The construction of 
$\widehat{K}'$ in the analogous
$2$-dimensional situation is illustrated in the figure below:

$$
\epsfxsize=13cm \epsfbox{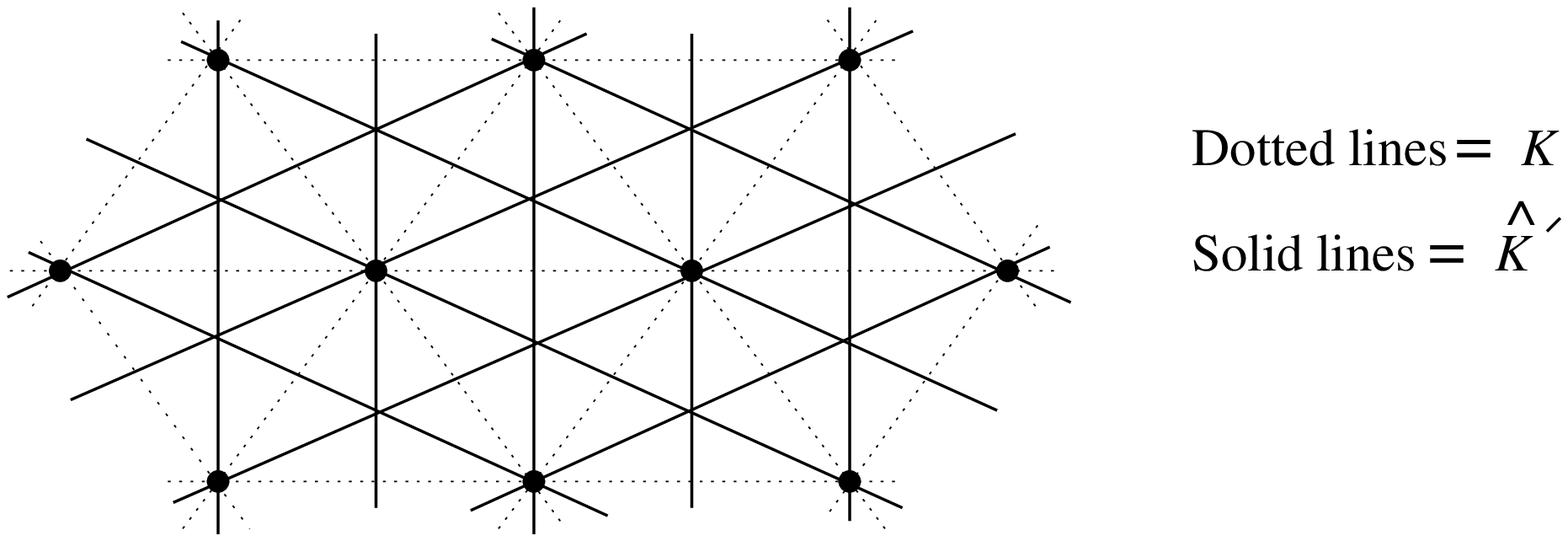}
$$

{\it Regularisation.}
The objects of interest in the simplicial theory can be expressed through
mathematically meaningful integrals 
over the finite-dimensional space $C^1(K)$ 
${\times}C^1(\widehat{K})$. These integrals are not 
convergent to start with however, because the simplicial
gauge invariance under ${\cal G}_K\times{\cal G}_{\widehat{K}}$ leads to
the appearance of divergent volumes of orbits of the simplicial gauge
group. Therefore we must proceed heuristically at first 
(as in the continuum case) and rewrite the
integrals as integrals over the simplicial orbit space
$C^1(K)/{\cal G}_K{\times}C^1(\widehat{K})/{\cal G}_{\widehat{K}}$. 
The resulting integrals for the partition function and Wilson v.e.v.'s 
in the simplicial theory are not convergent a priori, but become convergent
when we make the following {\it regularisation} of the simplicial theory:
Let $T$ be the selfadjoint map on $V\,\equiv\,C^1(K){\times}C^1(\widehat{K})$
given by
\begin{eqnarray}
i\lambda'\widetilde{S}_K(v)=i\lambda'<v,Tv>\qquad\qquad\,v=(x,y)
\label{5.r1}
\end{eqnarray}
(the explicit expression for $T$ is clear from (\ref{5.2})). Then there is
an orthogonal decomposition $V=V_+{\oplus}V_-{\oplus}V_0\ $, 
$T=T_+{\oplus}T_-{\oplus}T_0$ where $V_+\,$ $V_-\,$ and $V_0$ denote the 
subspaces of positive--, negative-- and zero--modes of $T\,$, and
$T_+\,$, $T_-\,$ and $T_0$ denote the restrictions of $T$ to these
spaces. For $\epsilon\in{\bf R_+}$ we define the {\it regularised} simplicial
action functional by
\begin{eqnarray}
i\lambda'\widetilde{S}_K^{(\epsilon)}(v):=<v'\,,((i\lambda'+\epsilon)T_+
\oplus(i\lambda'-\epsilon)T_-)v'>
\label{5.r2}
\end{eqnarray}
where $v'$ is the projection of $v{\in}V$ onto $V_+{\oplus}V_-$.
This is clearly invariant under ${\cal G}_K\times{\cal G}_{\widehat{K}}$
and coincides with $i\lambda'\widetilde{S}_K$ in the limit $\epsilon{\to}0$.
With this regularisation the expressions for the partition function and 
Wilson v.e.v.'s in the simplicial theory in terms of integrals over the
simplicial orbit space are convergent: Due to (\ref{5.2}), (\ref{5.6}) 
and the analogue of (\ref{5.6}) for $\widehat{K}$ these 
integrals are of the form 
\begin{eqnarray}
I_{\epsilon}=\int_{V_+{\oplus}V_-}{\cal D\/}v\,e^{i<w,v>}\,e^{-i\lambda'
\widetilde{S}_K^{(\epsilon)}(v)}
\label{5.r3}
\end{eqnarray}
(modulo overall factors; here $w{\in}V_+{\oplus}V_-$) which clearly
converges for $\epsilon>0$ due to (\ref{5.r2}).
In the limit where the regularisation is lifted we find
\begin{eqnarray}
\lim_{\epsilon{\to}0}I_{\epsilon}=
e^{-\frac{i\pi}{4}\eta(T)}e^{\frac{i}{4\lambda'}<w,T^{-1}w>}
\left(\frac{\lambda'}{\pi}\right)^{-\zeta(T)/2}|{\det}'(T)|^{-1/2}
\label{5.r4}
\end{eqnarray}
where $\eta(T)={\dim}V_+-{\dim}V_-$ and $\zeta(T)={\dim}V_++{\dim}V_-$.
In the following we evaluate the partition function and Wilson v.e.v.'s 
for the regularised theory given by (\ref{5.r2}). The regularisation is 
then lifted (i.e. the limit $\epsilon{\to}0$ is taken) in the final 
expressions.

{\it The partition function.}
Let $\Psi_K(C^1(K))$ and $\Psi_{\widehat{K}}(C^1(\widehat{K}))$ be equipped 
with the metrics determined by requiring that $\Psi_K$ and $\Psi_{\widehat{K}}$
(given by (\ref{5.0.1})--(\ref{5.0.2})) be isometries (where $C^1(K)$ and
$C^1(\widehat{K})$ have the metrics given by their canonical inner products).
The partition function is then the formal object 
\begin{eqnarray}
\widetilde{Z}_K(M,\lambda')^{(\epsilon)}&:=&\frac{1}{V({\cal G}_K\times
{\cal G}_{\widehat{K}})}\int_{\Psi_K(C^1(K))\times\Psi_{\widehat{K}}(C^1(
\widehat{K}))}{\cal D\/}A{\cal D\/}A'\,e^{-i\lambda'
\widetilde{S}^{(\epsilon)}(A,A')} \nonumber \\
&=&\frac{1}{V({\cal G}_K)V({\cal G}_{\widehat{K}})}\int_{C^1(K)/{\cal G}_K
\,\times\,C^1(\widehat{K})/{\cal G}_{\widehat{K}}}{\cal D\/}{\lbrack}x\rbrack
{\cal D\/}{\lbrack}y\rbrack\,V({\lbrack}x\rbrack)V({\lbrack}y\rbrack)\,
e^{-i\lambda'\widetilde{S}_K^{(\epsilon)}
({\lbrack}x\rbrack,{\lbrack}y\rbrack)} \nonumber\\
& &\label{5.8}
\end{eqnarray}
where we have implemented the regularisation described above.
Using the fact that ${\cal G}_K\,$, ${\cal G}_{\widehat{K}}$ and their 
actions on $C^1(K)\,$, $C^1(\widehat{K})$ have analogous properties to
${\cal G}$ and its action on ${\cal C}$ the divergent volumes
$V({\lbrack}x\rbrack)$ and $V({\lbrack}y\rbrack)$ can be formally evaluated
as in the continuum case. The divergent parts factor out as 
$V({\cal G}_K)$ and $V({\cal G}_{\widehat{K}})$ and cancel against the
overall normalisation factor in (\ref{5.8}), and the resulting expression for
the regularised partition function is finite.
Evaluating this expression using (\ref{5.2}) and then lifting the
regularisation (i.e. taking the limit $\epsilon{\to}0$) using (\ref{5.r4})
we find (compare with (\ref{4.6})):
\begin{eqnarray}
\lefteqn{\widetilde{Z}_K(M,\lambda')} \nonumber \\
&=&V({\cal F}_K/{\cal G}_K)|\det(\phi_0^K)||\det(\phi_1^K)|^{-1}
|{\det}'(d_0^K)||{\det}'(\frac{\lambda'}{\pi}\ast^Kd_1^K)|^{-1/2} \nonumber \\
& &\times\,
V({\cal F}_{\widehat{K}}/{\cal G}_{\widehat{K}})
|\det(\phi_0^{\widehat{K}})||\det(\phi_1^{\widehat{K}})|^{-1}
|{\det}'(d_0^{\widehat{K}})||{\det}'(\frac{\lambda'}{\pi}
\ast^{\widehat{K}}d_1^{\widehat{K}})|^{-1/2} \nonumber \\
& &\label{5.9}
\end{eqnarray}
where $V({\cal F}_K/{\cal G}_K)$ is determined by the volume element of 
$H^1(K,{\bf R})$ via (\ref{5.6.5}), and $V({\cal F}_{\widehat{K}}/{\cal G}_
{\widehat{K}})$ is determined analogously\footnote{As in the continuum case
the absence of a phase factor in (\ref{5.9}) is because the operator in the
quadratic form (\ref{5.2}) has symmetric spectrum (this is easily seen
using proposition 2.4) and therefore vanishing $\eta$.}.
Since the volume elements of
$H^1(K,{\bf R})$ and $H^1(\widehat{K},{\bf R})$ are determined by the 
volume elements of $H_{dR}^1(M)$ via the de Rham map, we see by
comparing (\ref{4.3.5}) and (\ref{5.6.5}) (and the analogue of (\ref{5.6.5})
for $\widehat{K}$) that the de Rham map induces isometries
\begin{eqnarray}
{\cal F}_K/{\cal G}_K\ \cong\ {\cal F}/{\cal G}\quad\quad,\qquad\qquad
{\cal F}_{\widehat{K}}/{\cal G}_{\widehat{K}}\ \cong {\cal F}/{\cal G}
\label{5.10}
\end{eqnarray}
so $V({\cal F}_K/{\cal G}_K)=V({\cal F}_{\widehat{K}}/{\cal G}_{\widehat{K}})=
V({\cal F}/{\cal G})$. Now using proposition 2.4 we find
\begin{eqnarray}
\widetilde{Z}_K(M,\lambda')
=\left(\frac{\lambda'}{\pi}\right)^{-{\dim}H^0+{\dim}H^1+N_0^K-N_1^K}
V({\cal F}/{\cal G})^2\,\tau_R(K,d^K)\,\frac{a_0^K}{a_1^K}
\label{5.11}
\end{eqnarray}
where $N_q^K$ denotes the number of $q$-simplices\footnote{In deriving the 
exponent for $\frac{\lambda'}{\pi}$ in (\ref{5.11}) we have used the fact 
that $N_0^K-N_1^K-N_0^{\widehat{K}}+N_1^{\widehat{K}}$ is the Euler
characteristic of $M\,$, which vanishes since ${\dim}M$ is odd.} of $K$ and
\begin{eqnarray}
\tau_R(K,d^K)=\prod_{q=0}^3\left(|\det(\phi_q^K)||{\det}'(d_q^K)|\right)^{
(-1)^q}
\label{5.12}
\end{eqnarray}
(with $|{\det}'(d_3^K)|{\equiv}1$) is the R--torsion of $(K,d^K)\,$, 
depending on the choices of volume elements for the spaces $H_{dR}^q(M)$ 
(via the induced volume elements for the spaces $H^q(K,{\bf R})$) but
independent of the choice of triangulation $K$ \cite{Reid,Franz,Milnor,RS}, 
and where for $q=0,1$ we have set
\begin{eqnarray}
a_q^K:=|\det(\phi_q^{\widehat{K}})||\det(\phi_{3-q}^K)|\,.
\label{5.13}
\end{eqnarray}
We claim that
\begin{eqnarray}
a_q^K=a_q
\label{5.14}
\end{eqnarray}
where $a_q$ is given by (\ref{4.8.5}); in particular $a_q^K$ is independent
of $K$. This is an application of proposition 4.2 of \cite{RS},
which states (as a special case) that
$<\omega,\tau>=<A_{dR}^K(\omega)\,,
(\ast^K)^{-1}A_{dR}^{\widehat{K}}(\ast\tau)>$
for $\omega\,,\tau\in{\cal H}^q(M)$. This can be rewritten as 
$<\omega,\tau>=<\omega\,,L_q\tau>$ where $L_q$ is the map
\begin{eqnarray}
{\cal H}^q(M)&\stackrel{\ast}{\longrightarrow}&
{\cal H}^{n-q}(M)\stackrel{\phi_{n-q}}{\longrightarrow}
H^{n-q}(M)\stackrel{A_{dR}^{\widehat{K}}}{\longrightarrow}
H^{n-q}(\widehat{K},{\bf R})
\stackrel{(\phi_{n-q}^{\widehat{K}})^{-1}}{\longrightarrow}
{\cal H}^{n-q}(\widehat{K}) 
\nonumber \\
&\stackrel{(\ast^K)^{-1}}{\longrightarrow}&
{\cal H}^q(K)\stackrel{((\phi_q^K)^{-1})^*}{\longrightarrow}
H^q(K,{\bf R})\stackrel{(A_q^K)^*}{\longrightarrow}
H^q(M)\stackrel{\phi_q^*}{\longrightarrow}{\cal H}^q(M)\,.
\label{5.14.5}
\end{eqnarray}
$L_q$ must coincide with the identity map, so we see from (\ref{5.14.5}) that
\begin{eqnarray*}
1=|\det(L_q)|=|\det(\phi_q)||\det(\phi_{n-q})||\det(\phi_{n-q}^{\widehat{K}})|^
{-1}|\det(\phi_q^K)|^{-1}
\end{eqnarray*}
and (\ref{5.14}) follows.

Since $\tau_R(K,d^K)=\tau_{RS}(M,d)$ \cite{Muller}
\cite{Cheeger} it follows that the final expression (\ref{5.11}) for
$\widetilde{Z}_K(M,{\lambda}')$ coincides with
the continuum expression $\widetilde{Z}(M,\lambda)=|Z(M,\lambda)|^2$ given by
the modulus square of (\ref{4.7}) except for the triangulation-dependent
factors $N_0^K$ and $N_1^K$ in the exponent of $\frac{{\lambda}'}{\pi}$.
This $K$-dependence can be removed by renormalising ${\lambda}'\,$, 
i.e. by replacing the bare coupling parameter ${\lambda}'$ by the 
renormalised $K$-dependent parameter ${\lambda}_K'$ given by
\begin{eqnarray}
\frac{{\lambda}_K'}{\pi}:=\left(\frac{{\lambda}'}{\pi}\right)^{\kappa(K)}
\quad\quad,\qquad\quad\kappa(K)=\left(1+\frac{N_0^K-N_1^K}{{\dim}H^1-
{\dim}H^0}\right)^{-1}
\label{5.16}
\end{eqnarray}
Then $\widetilde{Z}_K(M,{\lambda}_K')=\widetilde{Z}(M,\lambda)=
|Z(M,\lambda)|^2$.

{\it Wilson v.e.v.'s of framed loops.}
Let $\gamma_K^{(1)},\dots,\gamma_K^{(r)}$ be bounding edge loops in the 
triangulation $K\,$; i.e. each $\gamma_K^{(j)}$ is a 
bounding\footnote{Our considerations continue to hold in the more general
case where the $\gamma_K^{(j)}$'s represent torsion elements of the
homology of $M\,$, as discussed in \S6.} simplicial
map $\gamma_K^{(j)}:L_j{\to}K\,$, $S^1{\to}M$ (for some simplicial complex
$L_j$ triangulating $S^1$), and assume that the images $\gamma_K^{(j)}(S^1)$
are mutually disjoint in $M$. By theorem 3.6 and observation 3.8(iv) we
can choose a simplicial framing $\gamma^{(j)}$ for each $\gamma_K^{(j)}$
such that $\gamma^{(j)}$ and $\gamma_K^{(m)}$  
are mutually strongly disjoint for $j{\ne}m$.
Let $\gamma_{\widehat{K}}^{(j)}:\widehat{L_j'}{\to}\widehat{K}\,$, $S^1{\to}M$
denote the dual-simplicial map homotopic to $\gamma_K^{(j)}$ via
$\gamma^{(j)}$. Then the Wilson v.e.v. of the framed loops $\gamma^{(1)},
\dots,\gamma^{(r)}$ in the simplicial theory is given by the formal
expression
\begin{eqnarray}
\lefteqn{<\widetilde{W}_K(\gamma^{(1)},\dots,\gamma^{(r)};n_1,
\dots,n_r)>_{\lambda'}^{(\epsilon)}} \nonumber \\
&=&\frac{\int_{\Psi_K(C^1(K))\times\Psi_{\widehat{K}}(C^1(\widehat{K}))}
{\cal D\/}A{\cal D\/}A'\left\lbrack\prod_{j=1}^r\Phi(A,\gamma_K^{(j)},n_j)
\Phi(A',\gamma_{\widehat{K}}^{(j)},n_j)\right\rbrack\,e^{-i\lambda'
\widetilde{S}^{(\epsilon)}(A,A')}}{\int_{\Psi_K(C^1(K))\times
\Psi_{\widehat{K}}(C^1(\widehat{K}))}{\cal D\/}A{\cal D\/}A'\,e^{-i\lambda'
\widetilde{S}^{(\epsilon)}(A,A')}} \nonumber \\
& &\label{5.17}
\end{eqnarray}
where we have again implemented the regularisation described above.
For notational simplicity we set $\underline{\gamma}_K^{(j)}:=
\gamma_{K\#}^{(j)}({\lbrack}S^1\rbrack_{L_j}){\in}C_1(K)$ and
define $\underline{\gamma}_{\widehat{K}}^{(j)}{\in}C_1(\widehat{K})$
analogously.
The monodromies appearing in (\ref{5.17}) are
\begin{eqnarray}
\Phi(\Psi_K(x),\gamma_K^{(j)},n_j)&=&\Phi(A_0,\gamma_K^{(j)},n_j)\exp\left(
2{\pi}in_j\int_{S^1}\gamma_K^{(j)\ast}(W^{BK}(Bx))\right) \nonumber \\
&=&\Phi(A_0,\gamma_K^{(j)},n_j)\exp\left(
2{\pi}in_j<x\,,\underline{\gamma}_K^{(j)}>\right)
\label{5.18}
\end{eqnarray}
where the last equality follows from the definition of the map $B$ in \S2
and the property (\ref{2.3}) of $W^{BK}\,$, and similarly
\begin{eqnarray}
\Phi(\Psi_{\widehat{K}}(y),\gamma_{\widehat{K}}^{(j)},n_j)=
\Phi(A_0,\gamma_{\widehat{K}}^{(j)},n_j)\exp\left(
2{\pi}in_j<y\,,\underline{\gamma}_{\widehat{K}}^{(j)}>\right)
\label{5.19}
\end{eqnarray}
Note that if $A$ is a flat connection then $\Phi(A,\gamma_K^{(j)},n_j)=\Phi(A,
\gamma_{\widehat{K}}^{(j)},n_j)$ since $\gamma_K^{(j)}$ and 
$\gamma_{\widehat{K}}^{(j)}$ are homotopic.

The integrals in (\ref{5.17}) can be formally rewritten as integrals over 
the simplicial orbit space as previously. The divergent factors
$V({\cal G}_K)$ and $V({\cal G}_{\widehat{K}})$ in the
formal volumes of simplicial orbits which appear in the numerator and
denominator cancel, and the resulting expression for the 
regularised Wilson v.e.v. is finite.
Evaluating this expression using (\ref{5.2}), (\ref{5.10}), 
(\ref{5.r4}) and proposition 2.4 and lifting
the regularisation we find (compare with (\ref{4.14})):
\begin{eqnarray}
\lefteqn{<\widetilde{W}_K(\gamma^{(0)},\dots,\gamma^{(r)};n_1,
\dots,n_r)>_{\lambda'}} \nonumber \\
&=&\left(\frac{1}{V({\cal F}/{\cal G})}\int_{{\cal F}/{\cal G}}{\cal D\/}
{\lbrack}A\rbrack\,\prod_{l=1}^r\Phi({\lbrack}A\rbrack,\gamma_K^{(l)},n_l)
\right)^2 \nonumber \\
& &\times\,\exp\left(\frac{2i\pi^2}{\lambda'}\sum_{j\,m}n_jn_m
<\underline{\gamma}_K^{(j)}\,,(\ast^Kd^K)^{-1}
\underline{\gamma}_{\widehat{K}}^{(m)}>\right)
\label{5.20}
\end{eqnarray}
By theorem 3.3 and proposition 3.9 
the quantity $<\underline{\gamma}_K^{(j)}\,,(\ast^Kd^K)^{-1}
\underline{\gamma}_{\widehat{K}}^{(m)}>$ in (\ref{5.20}) equals
the linking number of $\gamma_K^{(j)}$ and $\gamma_K^{(m)}$ (or the 
self-linking number if $j=m\,$), 
and it follows that (\ref{5.20}) coincides with the 
continuum expression (\ref{4.14}) for the v.e.v. (No renormalisation of 
$\lambda'$ is required in this case.)
This equality continues to hold when
$\gamma_K^{(1)},\dots,\gamma_K^{(r)}$ represent torsion elements
of the ${\bf Z}-$homology of $K\,$, as discussed in the next section.

The techniques of this section generalise in a straightforward way to 
construct simplicial versions of the abelian $BF$ theories associated
with the generalisations of abelian Chern--Simons theory discussed at the
end of \S4, again reproducing the continuum expressions for the 
partition function and Wilson v.e.v.'s of generalised, simplicially framed 
edge loops, expressed in terms of R--torsion and linking numbers
respectively. We omit the details, except to point out that in these
theories the simplicial gauge symmetry is simpler than above:
The continuum gauge transformations $(\omega\,,\omega')\mapsto(\omega+d\tau\,,
\omega'+d\tau')$ becomes $(x,y)\mapsto(x+d^Ku\,,y+d^{\widehat{K}}v)$ in the
simplicial theory.

\section{Torsion pairings from Wilson v.e.v.'s at discrete values of the
coupling parameter}

In this section we consider the Wilson
v.e.v.'s in the case where the loops
are no longer required to be bounding but instead represent torsion elements
of the ${\bf Z}$--homology of $M$. We will see that in this case 
${\bf Q}/{\bf Z}$--valued torsion pairings appear in place of linking numbers
in the v.e.v.'s. when the coupling parameter takes the discrete values 
$\lambda=\frac{\pi}{2l}\,$, $l\in{\bf Z}$. 
We begin by briefly reviewing torsion pairings.
(A brief accessible review of torsion in (co)homology with illuminating
examples is given in \S3 of \cite{Freed}. More detailed treatments can
be found in \cite{Novikov} \cite{BottTu} \cite{Munkres}.)

{\it Torsion pairings}. Let $M$ be a closed oriented $n$-manifold with
triangulation $K$ as in \S2--3. 
$M$ need not be smooth, and all maps in the following are continuous.
An element $x$ of the (singular) 
${\bf Z}$-homology of $M\,$, $H_*(M,{\bf Z})\,$, 
is said to be a torsion element if $kx=0$ for some non-zero $k\in{\bf Z}$. 
These elements form a subgroup $\mbox{Tor}H_*(M,{\bf Z})$.
Torsion elements can arise as follows. Let $N$ be a closed oriented
$p$-manifold, then the element 
$\lbrack\underline{f}\rbrack{\in}H_p(M,{\bf Z})$ represented by a 
map $f:N{\to}M$ is a torsion element if there is a map
$h:D{\to}M$ where $D$ is an oriented $(p+1)$-manifold with boundary 
${\partial}D$ consisting of $k$ disjoint copies of $N$ and the restriction
of $h$ to ${\partial}D$ is $f$ on each copy, since in this case 
${\partial}\underline{h}=k\underline{f}$ where 
$\underline{f}$ and $\underline{h}$
are the chains over ${\bf Z}$ corresponding to $f$ and $h$ respectively.

A standard result concerning torsion (see e.g. \S3 of \cite{Freed}) is that
there is a canonical isomorphism
\begin{eqnarray}
\mbox{Tor}H^q(M,{\bf Z})\;\simeq\;\mbox{Hom}(\mbox{Tor}H_{q-1}(M,{\bf Z})
\,,\,{\bf Q}/{\bf Z})\,. 
\label{6.1}
\end{eqnarray}
defined as follows. Let $x$ be a $(q-1)$-cycle over ${\bf Z}$ representing
an element ${\lbrack}x\rbrack\in\mbox{Tor}_{q-1}(M,{\bf Z})\,$, then there
is a non-zero integer $k$ and a $q$-chain $y$ over ${\bf Z}$ such that
${\partial}y=kx$. Let $c$ be a $q$-cocycle over ${\bf Z}$ representing an
element ${\lbrack}c\rbrack\in\mbox{Tor}H^q(M,{\bf Z})\,$, then the pairing
(\ref{6.1}) is given by
\begin{eqnarray}
\langle{\lbrack}c\rbrack\,{\lbrack}x\rbrack\rangle:=\frac{1}{k}<c\,,y>
\qquad(\;\mbox{mod}\;{\bf Z})
\label{6.1.5}
\end{eqnarray}
It is easy to check that $\frac{1}{k}<c\,,y>$ depends only on 
${\lbrack}c\rbrack$ and ${\lbrack}x\rbrack$ (mod ${\bf Z}$) as required.
Poincare duality gives $\mbox{Tor}H^q(M,{\bf Z})\,\simeq\,
\mbox{Tor}H_{n-q}(M,{\bf Z})\,$,
allowing (\ref{6.1}) to be re-expressed (with $p=n-q$) as
\begin{eqnarray}
\mbox{Tor}H_p(M,{\bf Z})\;\simeq\;\mbox{Hom}(\mbox{Tor}H_{n-p-1}(M,{\bf Z})
\,,\,{\bf Q}/{\bf Z})\,. 
\label{6.2}
\end{eqnarray}
Thus there is a canonical ${\bf Q}/{\bf Z}$--valued pairing between 
$\mbox{Tor}H_p(M,{\bf Z})$ and $\mbox{Tor}H_{n-p-1}(M,{\bf Z})\,$, 
which we call the torsion
pairing. This pairing arises as a direct generalisation of linking number
as follows. Let $N_1$ and $N_2$ be closed oriented manifolds of dimensions
$p$ and $n-p-1$ respectively, and for $i=1,2$ let $f_i:N_i{\to}M$ and 
$h_i:D_i{\to}M$ be analogous to $f:N{\to}M$ and $h:D{\to}M$ above, so
that ${\partial}\underline{h}_i=k_i\underline{f}_i$ at the chain level and 
${\lbrack}\underline{f}_1\rbrack{\in}\mbox{Tor}H_p(M,{\bf Z})\,$, 
${\lbrack}\underline{f}_2\rbrack{\in}\mbox{Tor}H_{n-p-1}(M,{\bf Z})$.
The chain $k_i\underline{f}_i$ corresponds to the map,
denoted $k_if_i\,$, from $k_i$ copies of $N_i$ into $M$ coinciding
with $f_i$ on each copy. The maps $k_1f_1$ and $k_2f_2$ are bounding
and therefore have well-defined linking number.

{\it Proposition 6.1}. 
The torsion pairing (\ref{6.2}) of ${\lbrack}\underline{f}_1\rbrack$ and 
${\lbrack}\underline{f}_2\rbrack$ above is given by
\begin{eqnarray}
\mbox{tor}({\lbrack}\underline{f}_1\rbrack\,,{\lbrack}\underline{f}_2\rbrack)=
\frac{1}{k_1k_2}\mbox{lk}(k_1f_1\,,k_2f_2)\qquad(\;\mbox{mod}\;{\bf Z})
\label{6.3}
\end{eqnarray}
This follows via simplicial approximation from its simplicial version,
proposition 6.1$\,'$ below.

The considerations above have analogues in the simplicial setting. 
The duality operator (simplicial Hodge star operator) $\ast^K:C^q(K){\to}
C^{n-q}(\widehat{K}){\cong}C_{n-q}(\widehat{K})$ (where the (co)chains are
over ${\bf R}$) restricts to $\ast^K:C^q(K,{\bf Z}){\to}C_{n-q}(\widehat{K},
{\bf Z})\,$, inducing isomorphisms (Poincare dualities) 
\begin{eqnarray*}
\ast^K:H^q(K,{\bf Z})\stackrel{\cong}{\to}H_{n-q}(\widehat{K},{\bf Z})
\qquad,\qquad\ast^K:\mbox{Tor}H^q(K,{\bf Z})\stackrel{\cong}{\to}
\mbox{Tor}H_{n-q}(\widehat{K},{\bf Z})
\end{eqnarray*}
(These are standard facts which follow easily from proposition 2.4).
Thus the simplicial version of (\ref{6.2}) is 
\begin{eqnarray}
\mbox{Tor}H_p(K,{\bf Z})\;\simeq\;\mbox{Hom}(\mbox{Tor}H_{n-p-1}
(\widehat{K},{\bf Z})\,,\,{\bf Q}/{\bf Z})
\label{6.5}
\end{eqnarray}
i.e. the ${\bf Q}/{\bf Z}$--valued torsion pairing is between 
$\mbox{Tor}H^p(K,{\bf Z})$ and $\mbox{Tor}H_{n-p}(\widehat{K},{\bf Z})$.
Let $f_K:L_1{\to}K\,$, $N_1{\to}M$ and $g_{\widehat{K}}:\widehat{L_2}{\to}K\,$,
$N_2{\to}M$ be simplicial- and dual-simplicial maps respectively as in \S3,
and let $h_1:D_1{\to}M$ and $h_2:D_2{\to}M$ be maps with
${\partial}D_i$ consisting of $k_i$ copies of $N_i$ ($i=1,2$) such that
$h_1$ restricts to $f_K$ and $h_2$ restricts to $g_{\widehat{K}}$ on each
copy of $N_1$ and $N_2$ respectively. Let $\underline{f}_K{\in}C_p(K,{\bf Z})$
and $\underline{g}_{\widehat{K}}{\in}C_{n-p-1}(\widehat{K},{\bf Z})$ denote
the chains corresponding to $f_K$ and $g_{\widehat{K}}$.
By the arguments of \S3 (see lemmas 3.2 and 3.4)
we can assume that $h_1$ and $h_2$ determine elements
$\underline{h}_1{\in}C_{p+1}(K,{\bf Z})\,$, $\underline{h}_2{\in}C_{n-p-1}
(\widehat{K},{\bf Z})$ with 
${\partial}^K\underline{h}_1=k_1\underline{f}_K$ and 
${\partial}^{\widehat{K}}\underline{h}_2=k_2\underline{g}_{\widehat{K}}$. 
Then $\lbrack\underline{f}_K\rbrack\in\mbox{Tor}H_p(K,{\bf Z})$ and 
$\lbrack\underline{g}_{\widehat{K}}\rbrack\in\mbox{Tor}H_{n-p-1}(\widehat{K},
{\bf Z})$.

{\it Proposition 6.1$\,'$}.
The torsion pairing (\ref{6.5}) of $\lbrack\underline{f}_K\rbrack$ and 
$\lbrack\underline{g}_{\widehat{K}}\rbrack$ above is given by
\begin{eqnarray}
\mbox{tor}(
{\lbrack}\underline{f}_K\rbrack\,,
{\lbrack}\underline{g}_{\widehat{K}}\rbrack)=
\frac{1}{k_1k_2}\mbox{lk}(k_1f_K,k_2g_{\widehat{K}})
\label{6.6}
\end{eqnarray}
where $k_1f_K$ is the map from $k_1$ copies of $N_1$ into $M$ coinciding
with $f_K$ on each copy, and $k_2g_{\widehat{K}}$ is defined analogously.

\noindent {\it Proof}.
By (\ref{6.1.5}),
\begin{eqnarray*}
\mbox{tor}(\lbrack\underline{f}_K\rbrack\,,\lbrack\underline{g}_{\widehat{K}}
\rbrack)\ \equiv\ \langle(\ast^{\widehat{K}})^{-1}
\lbrack\underline{f}_K\rbrack\,,\lbrack\underline{g}_{\widehat{K}}
\rbrack\rangle=\frac{1}{k_2}<(\ast^{\widehat{K}})^{-1}\underline{f}_K\,,
\underline{h_2}>
\end{eqnarray*}
(mod ${\bf Z}$). Using proposition 2.4 this can be rewritten as 
\begin{eqnarray*}
\frac{1}{k_1k_2}<k_1\underline{f}_K\,,
(\ast^Kd^K)^{-1}(k_2\underline{g}_{\widehat{K}})>
\qquad(\;\mbox{mod}\;{\bf Z}\;)
\end{eqnarray*}
and the proposition follows from theorem 3.3. \qed

{\it The v.e.v.'s of loops representing torsion elements.}
The evaluations of the v.e.v.'s of loops $\gamma^{(1)},\dots,\gamma^{(r)}$ 
in \S4--5 goes through with obvious modifications in the case where the
loops represent torsion elements of the ${\bf Z}$--homology of $M\,$:
Choose numbers $k_j$ so that $k_j\gamma^{(j)}$ is bounding, then after
dividing by $k_j$ in the appropriate places we can replace $\gamma^{(j)}$
by $k_j\gamma^{(j)}$ in the evaluations of the v.e.v.'s. This results in
the linking numbers $\mbox{lk}(\gamma^{(j)},\gamma^{(m)})$ being replaced
by $\frac{1}{k_jk_m}\mbox{lk}(k_j\gamma^{(j)},k_m\gamma^{(m)})$ in the final
expressions (\ref{4.10}), (\ref{4.14}) and (\ref{5.20}) for the v.e.v.'s.
This now results in non-trivial expressions when the coupling parameter 
takes the discrete values
$\lambda=\frac{\pi}{2l}\,$, $l\in{\bf Z}$ in the abelian 
Chern--Simons theory, or as $\lambda=\frac{\pi}{l}$ ($\lambda'=\frac{\pi}{l}$)
in the corresponding abelian BF theory (and its simplicial version).
It follows from (\ref{6.3}) and (\ref{6.6}) that in this case the v.e.v.'s
only depend on the homology classes of the loops, and that the linking numbers
get replaced by the 
${\bf Q}/{\bf Z}-$valued torsion pairings
$\mbox{tor}(\lbrack\underline{\gamma}^{(j)}\rbrack,
\lbrack\underline{\gamma}^{(m)}\rbrack)$ in the final expressions for the 
v.e.v.'s. (The framing considerations go through as before.)

There is one other modification in the final expressions for the v.e.v.'s
when the loops represent torsion elements. It comes from the integral
over the modulispace ${\cal F}/{\cal G}$ in the expression for the v.e.v.'s
(see (\ref{4.10})), due to the following

{\it Observation/definition 6.2}.
If $\gamma:S^1{\to}M$ represents a torsion element, i.e. if there is some
map $h:D{\to}M$ with ${\partial}D=kS^1$ (i.e. $k$ copies of $S^1$) with
$h\Big|_{{\partial}D}=k\gamma$ then the monodromy $\Phi(A,\gamma,n)$
is the same for all flat connections $A$. 
For given $\gamma$ and $n$ this monodromy therefore 
depends only on the underlying flat principal $U(1)$ bundle $P\,$, 
and we denote it by $\Phi(P,\gamma,n)$.

\noindent To see this let $A$ and $A'$ be flat connections, then
$A'=A+2{\pi}i\omega$ with $\omega\in\ker(d_1)$ and we have
\begin{eqnarray*}
\int_{S^1}\gamma^*(\omega)=\frac{1}{k}\int_{{\partial}D=kS^1}(k\gamma)^*
(\omega)=\frac{1}{k}\int_Dh^*(d\omega)=0
\end{eqnarray*}
which implies $\Phi(A',\gamma,n)=\Phi(A,\gamma,n)$.
(Note also that ${\partial}h=k\gamma$ implies $\Phi(P,\gamma,n)^k=1$.)
Thus the integral over ${\cal F}/{\cal G}$ in 
(\ref{4.10}) leads in the present case to an overall factor
$\prod_{j=1}^r\Phi(P,\gamma^{(j)},n_j)$ appearing in the final expression
(\ref{4.10}), while
the square of this factor appears in the final expressions (\ref{4.14})
and (\ref{5.20}).

These considerations also go through in the previously mentioned
generalisations of abelian Chern--Simons theory and in the corresponding
BF theories.

As discussed in \cite{Freed}, torsion phenomena arise in connection with
global anomalies in string theory. This suggests a connection between
anomalies and abelian Chern--Simons theory on 3-dimensional parameter
spaces of families of Dirac operators. We will be discussing this connection
in a forthcoming paper.

{\it Acknowledgements.} 
I am grateful to Prof. Siddhartha Sen for many 
helpful discussions and encouragement
during the course of this work. I also thank Dr. Jim 
Sexton for pointing out the analogy with lattice fermion doubling, 
and Dr. Denjoe O'Conner for drawing
my attention to some important background references. The main results of
this paper were first presented at the 3rd Irish QFT Meeting, Dublin,
May '96, and I thank Prof.'s A.P.~Balachandran and G.~Segal for their interest
on that occasion. This work was partially supported by FORBAIRT.
I thank the Dublin Institute for Advanced Studies for hospitality and 
financial support during the summer '96.

\end{document}